\documentclass[11pt]{article}
\usepackage{putex}
\usepackage{graphicx}

\begin{document}
\preprint{COLO-HEP-575 \\ PUPT-2422}

\institution{CU}{${}^1$Department of Physics, 390 UCB, University of Colorado, Boulder, CO 80309, USA}
\institution{PU}{${}^2$Joseph Henry Laboratories, Princeton University, Princeton, NJ 08544, USA}

\title{Fermi surfaces in ${\cal N}=4$ Super-Yang-Mills theory}

\authors{Oliver DeWolfe,${}^\CU$ Steven S.~Gubser,${}^\PU$ and Christopher Rosen${}^\CU$}

\abstract{We investigate and classify Fermi surface behavior for a set of fermionic modes in a family of backgrounds holographically dual to ${\cal N}=4$ Super-Yang-Mills theory at zero temperature with two distinct chemical potentials.  We numerically solve fluctuation equations for every spin-1/2 field in  five-dimensional maximally supersymmetric gauged supergravity not mixing with gravitini.  Different modes manifest two, one or zero Fermi surface singularities, all associated to non-Fermi liquids, and we calculate dispersion relations  and widths of excitations.  We study two limits where the zero-temperature entropy vanishes.
In one limit, a Fermi surface approaches a marginal Fermi liquid, which we demonstrate analytically, and conductivity calculations show a hard gap with the current dual to the active gauge field superconducting, while the other is insulating.  In the other limit, conductivities reveal a soft gap with the roles of the gauge fields reversed.
}

\date{July 2012}

\maketitle

\section{Introduction}

States of ordinary metals are well-described by Fermi liquid theory, consisting of weakly-coupled, long-lived quasiparticle excitations around a Fermi surface.  It is of interest, however, to move beyond standard Fermi liquid theory; indeed, certain strongly correlated electron systems, such as 
the ``strange metals" arising in high-$T_c$ cuprates \cite{Varma, Anderson} and heavy fermion systems \cite{Gegenwart}, are known to possess a Fermi surface from photoemission experiments, but the associated gapless excitations are not long-lived.  Such systems, with Fermi surfaces but without ordinary quasiparticles, have been called  ``non-Fermi liquids", and developing frameworks to characterize them is important from the theoretical point of view.

The gauge-gravity correspondence  \cite{Maldacena:1997re,Gubser:1998bc,Witten:1998qj} is a promising avenue to investigate these phenomena, as strongly-coupled systems which are difficult to characterize in non-gravitational variables are described holographically by black hole geometries that may be far more tractable calculationally. Holographic Fermi surfaces were first investigated in \cite{Lee:2008xf,Liu:2009dm, Cubrovic:2009ye}, and have been the subject of substantial study.  
Non-Fermi liquid systems have indeed been obtained;
 a study of Fermi surface behavior for general dimension, mass and charges was carried out in \cite{Faulkner:2009wj}.

Much of the work so far has followed a ``bottom-up" approach: a convenient effective Lagrangian in the gravity theory is postulated, and its consequences worked out.  However, while such an approach is quite valuable, it does not make direct contract with supergravity or string theory.  The potential drawbacks are twofold.  First, the exact nature of the dual field theory remains unknown, as only the symmetries can be reliably established without the string theory map, so one is
ignorant of exactly what system is supporting the Fermi surface.  Second, without the guidance of a known string theory construction, one may worry that the masses or charges selected may contain a hidden instability or other secret unphysical property, and therefore that the results may be less trustworthy.
Therefore, a ``top-down" approach, where one starts from a known string or supergravity construction, is naturally valuable both for understanding the system and having confidence in its validity.
 Previous top-down approaches to Fermi surfaces include \cite{Ammon:2010pg, Jensen:2011su}, which studied fermions realized on probe branes, and \cite{Gauntlett:2011mf, Belliard:2011qq, Gauntlett:2011wm}, which studied the gravitino in the gravity multiplet of minimal supergravity and found no Fermi surface singularity; \cite{Berkooz:2012qh} (see also \cite{Berkooz:2006wc, Berkooz:2008gc}) discusses rotating black holes with zero entropy in supergravity and relates them to Fermi surfaces from a different perspective.  For other studies of finite density systems using the gauge-gravity correspondence, see for example \cite{Denef:2009yy}-
\cite{Gopakumar:2012gd}; recent reviews appear in \cite{Hartnoll:2011fn, 2011arXiv1108.1197S, Iqbal:2011ae}.
 
Many known holographic examples also have a nonzero entropy remaining even at zero temperature.  This characteristic is somewhat mysterious physically and is unlike most strongly correlated systems that one might want to apply the gauge/gravity correspondence to.  Thus, it is also useful to pursue examples of holographic Fermi surfaces where the zero-temperature entropy vanishes.

In \cite{DeWolfe:2011aa}, we presented top-down constructions of Fermi surfaces in five- and four-dimensional maximally supersymmetric gauged supergravity, dual to four-dimensional ${\cal N}=4$ Super-Yang-Mills (SYM) theory and three-dimensional ABJM theory, respectively.  The work \cite{DeWolfe:2011aa} studied for each case a single mode in a single background corresponding to all chemical potentials equal, and was primarily concerned with establishing the existence of top-down holographic Fermi surfaces.  It is natural, however, to wish to go beyond a single case to study a wide variety of fermionic modes, in a variety of backgrounds with varying chemical potential, and attempt to classify the diversity of Fermi surface behavior. Moreover, the explicit knowledge of the dual field theory in top-down constructions means an understanding of the dual zero-temperature states is within reach, and understanding a broader class of solutions can only help make this identification.  In this paper, we undertake such a study for the case of ${\cal N}=4$ Super-Yang-Mills.

${\cal N}=4$ Super-Yang-Mills has an $SO(6)$ R-symmetry within which three distinct chemical potentials may be turned on.  Here we study the class of extremal black brane geometries dual to zero-temperature states in which two of these chemical potentials are set equal to one another, while the third may vary separately; the dimensionless ratio provides us with a one-parameter family of geometries.  The cases where one chemical potential or the other is set to zero are also of interest, and we study extremal limits of these as well: one is a non-thermodynamic BPS state on the Coulomb branch, and the other is a black hole with vanishing entropy at zero temperature.  We study all the spin-1/2 fermionic modes of five-dimensional maximal gauged supergravity --- corresponding to the smallest-dimension fermionic operators in the large-$N$, large-coupling limit --- that do not mix with the spin-3/2 gravitino fields.  There are 18 such modes, along with their conjugates, obeying 8 distinct Dirac equations.  Characteristic of these equations are scalar-dependent (and therefore position-dependent) mass and Pauli terms, absent in the usual class of bottom-up models.

The extremal geometries with two distinct chemical potentials, called the 2+1-charge black holes, 
are of a class with a double pole at the horizon studied in general in \cite{Faulkner:2009wj}.  In this paper we numerically solve the various Dirac equations in these backgrounds, looking for Fermi surface singularities. We find that fermions dual to Tr $F_+ \lambda$ operators made from a field strength and a gaugino have no Fermi surfaces, while fermions dual to Tr $\lambda X$ made from a gaugino and a scalar may have two, one or zero Fermi surfaces, depending on the net charge.  Double fermi surfaces appear as nested spheres in $k$-space.  These systems all behave as non-Fermi liquids, with excitations that are not stable in the low-energy  limit and hence are not true quasiparticles; we determine dispersion relations and width per energy ratios for these excitations.  The fermions also generically possess so-called oscillatory regions manifesting log-periodic behavior where the Fermi surfaces disappear \cite{Liu:2009dm, Faulkner:2009wj}, and we identify these as well.

There are two interesting limits of our family associated to vanishing zero-temperature entropy.  One is associated to taking the two identical chemical potentials much larger than the third, and we show that in this limit one of our Fermi surface singularities approaches a marginal Fermi liquid (MFL), which occupies the border between Fermi and non-Fermi liquids, and was postulated to describe the optimally doped cuprates \cite{Varma}.   We are able to show analytically that this behavior is exact, and is controlled by the extremal limit of the ``1-charge" black hole where the two equal chemical potentials strictly vanish, which is a non-thermodynamic BPS Coulomb branch state \cite{Freedman:1999gk}.  We solve the fermion fluctuation spectrum in this background, completing the analysis of \cite{Bianchi:2000sm}.  We calculate conductivities in this geometry as well, and find a hard gap manifested by a step function, and argue that the current dual to the active gauge field is superconducting, while the other is insulating.  We also determine the conductivities in the 2+1-charge backgrounds near the marginal Fermi liquid limit, which we may think of as ``doping" the 1-charge background with a little of the other charge, and see that the delta function in the conductivity associated to superconducting behavior in the 1-charge background broadens into a Drude peak.

The extremal limit of the ``2-charge" black hole where the identical chemical potentials are nonzero but the other is zero is also of interest, for it is also dual to a system with zero entropy at zero temperature.  We study the fermion fluctuation equations in this geometry as well, and find that while two modes have Fermi surface singularities, the form of the nearby small energy fluctuations is damped rather than infalling.  This system displays a soft gap with a rapid but smooth rise in conductivity, again with a zero-frequency delta function in the active gauge field, although this may be a result of translation invariance rather than superconductivity.

In section~\ref{BlackHolesSec} we review the charged black hole solutions we will use, as well as their field theory duals, and in section~\ref{GaugedSUGRASec} we show how these solutions are embedded within five-dimensional maximally supersymmetric gauged supergravity.  In section~\ref{FermionSec} we determine the Dirac equations in these backgrounds for the spin-1/2 fields of gauged supergravity, as well as describing their dual operators.  In section~\ref{FermiDiracSec} we discuss the near-horizon behavior of the Dirac equations and make general statements about oscillatory regions and Fermi surfaces, and
in section~\ref{TwoPlusOneSec} we solve these Dirac equations in the 2+1-charge backgrounds and obtain Fermi surfaces, whose properties we plot as functions of the ratio of chemical potentials.  Section~\ref{OneChargeSec} studies the 1-charge black hole, enabling us to analytically derive the marginal Fermi liquid limit, as well as calculating the conductivities in the 1-charge and nearby 2+1-charge backgrounds.  In section~\ref{TwoChargeSec} we determine Fermi surface singularities and conductivities in the 2-charge background, and in section~\ref{ConclusionsSec} we conclude.

\section{Charged black hole solutions}
\label{BlackHolesSec}

${\cal N}=4$ Super-Yang-Mills theory with $SU(N)$ gauge group is dual to type IIB string theory on $AdS_5 \times S^5$.  In the limit of large $N$ and large 't Hooft coupling, string theory may be approximated by type IIB supergravity.
The Kaluza-Klein reduction of type IIB supergravity on the five-sphere may be consistently truncated to only the lowest-mass modes constituting a single five-dimensional multiplet, resulting in five-dimensional maximally supersymmetric gauged supergravity. Each mode in the gauged supergravity theory is dual to a known operator in the field theory.

The reduction on the $S^5$ leads to an $SO(6)$ gauge group in five-dimensions.  Black brane solutions are known that are charged under the $SO(2) \times SO(2) \times SO(2) = U(1)_a \times U(1)_b \times U(1)_c$ Cartan subgroup.  (We will use the terms ``black brane" and ``black hole" interchangably; all our geometries have three-dimensional spatial translation invariance.)  The gauged supergravity theory also contains 42 scalar fields; general black holes with three unequal charges $(Q_a, Q_b, Q_c)$ source two distinct scalars and leave only the Cartan subgroup unbroken.  The consistent truncation of the full theory to include the metric, the three gauge fields and the two scalars is called the STU model \cite{Behrndt:1998jd}.

A simplification of this situation, which we pursue here, is to take two of the three charges equal; we take $Q_b = Q_c$ and define $Q_1 \equiv Q_a$, $Q_2 \equiv Q_b = Q_c$.  In this case we have a larger preserved symmetry, $SO(6) \to SO(2) \times SO(4)$, and only one scalar field is turned on.  
The effective Lagrangian for the metric, two gauge fields and single scalar (in mostly minus signature) is
\eqn{Lag}{
e^{-1} {\cal L} &=   -R + {1 \over 2} (\partial \phi)^2 + {8 \over L^2} e^{\phi \over \sqrt{6}} + {4 \over L^2} e^{-2 \phi \over \sqrt{6}}
- e^{-4 \phi \over \sqrt{6}} f_{\mu\nu} f^{\mu\nu}  - 2
e^{2 \phi \over \sqrt{6}} F_{\mu\nu} F^{\mu\nu}  
- 2 \epsilon^{\mu\nu\rho\sigma\tau} f_{\mu\nu} F_{\rho\sigma} A_\tau \,.
}
The unconventional gauge field normalizations will be convenient for matching to maximal gauged supergravity.
Here we will study a class of black brane solutions
\eqn{Background}{
ds^2 &= e^{2A(r)} (h(r) dt^2 - d \vec{x}^2) - {e^{2B(r)} \over h(r)} dr^2 \,, \cr
a_\mu dx^\mu &= \Phi_1(r)\, dt \,, \quad \quad A_\mu dx^\mu = \Phi_2(r)\, dt \,, \quad \quad \phi = \phi(r) \,.
}
Particular limits that have been studied and will play central roles here include the ``1-charge black hole" ($Q_2 = 0$), the ``2-charge black hole" ($Q_1 = 0$) and the ``3-charge black hole" ($Q_1 = Q_2$).  We will refer to a solution with general $Q_1$, $Q_2$ as a 2+1-charge black hole.  We now review these solutions.

\subsection{2+1-charge black holes and 3-charge case}

The 2+1-charge black hole involves both gauge fields $a_\mu$ and $A_\mu$ as well as the active scalar.  It satisfies the ansatz \eno{Background} with solution
\eqn{TwoPlusOneSoln}{
A(r) &= \log {r \over L} + {1 \over 6} \log \left( 1 + {Q_1^2\over r^2} \right) + {1 \over 3} \log \left( 1 + {Q_2^2\over r^2} \right)\,, \cr
B(r) &= - \log {r \over L}- {1 \over 3} \log \left( 1 + {Q_1^2\over r^2} \right)- {2 \over 3} \log \left( 1 + {Q_2^2\over r^2} \right)\,, \cr
h(r) &= 1- { r^2(r_H^2 + Q_1^2)(r_H^2 + Q_2^2)^2 \over r_H^2 (r^2 + Q_1^2)(r^2 + Q_2^2)^2 }\,, \quad
\phi(r) =  -\sqrt{2\over 3} \log \left( 1 + {Q_1^2\over r^2} \right) +\sqrt{2\over 3} \log \left( 1 + {Q_2^2\over r^2} \right) \,, \cr
\Phi_1(r) &={Q_1 (r_H^2 + Q_2^2) \over 2 L r_H\sqrt{r_H^2 + Q_1^2}} \left(1-{r_H^2 + Q_1^2 \over r^2 + Q_1^2}   \right) \,, \quad
\Phi_2(r) ={Q_2 \sqrt{r_H^2 + Q_1^2} \over 2 L r_H} \left(1-{r_H^2 + Q_2^2 \over r^2 + Q_2^2} \right) \,,
}
where we have traded in a mass parameter for the horizon radius $r_H$, which is the largest solution of $h(r_H) = 0$. In the limit of large $r$, the solutions approach anti-de Sitter space with radius $L$.
The temperature and chemical potentials are determined by the usual methods to be\footnote{The chemical potentials are defined relative to canonically normalized gauge fields.}
\eqn{}{
T = {2 r_H^4 + Q_1^2 r_H^2 - Q_1^2 Q_2^2 \over 2 \pi L^2 r_H^2 \sqrt{r_H^2 + Q_1^2}} \,, \quad
\mu_1 = {Q_1 (r_H^2 + Q_2^2) \over L^2 r_H \sqrt{r_H^2 + Q_1^2}}\,, \quad
\mu_2 =  {\sqrt{2} Q_2 \sqrt{r_H^2 + Q_1^2} \over L^2 r_H} \,,
}
and the entropy and charge densities are
\eqn{}{
s = {1 \over 4 G L^3} (r_H^2 + Q_1^2)^{1/2} (r_H^2 + Q_2^2) \,, \quad \quad
\rho_1 = {Q_1 s \over 2 \pi r_H} \,, \quad \quad \rho_2 = {\sqrt{2} Q_2 s \over 2 \pi r_H} \,.
}
An extremal black hole exists for any solution setting the temperature to zero with chemical potentials nonzero.  This requires a non-negative solution to 
\eqn{ExtremalRule}{
r_H^2 = {1 \over 4} \sqrt{Q_1^4 + 8 Q_1^2 Q_2^2} - {1 \over 4} Q_1^2 \,,
}
and it is easy to see for nonzero $Q_1$ and $Q_2$ such a solution always exists. Thus for general charges in the 2+1-charge black hole there is an extremal limit.  The entropy density is generally nonzero at extremality.  In practice it will be more convenient to solve the extremality condition \eno{ExtremalRule} by eliminating $Q_2$,
\eqn{Q2ExtremalRule}{
Q_2^2 = {2 r_H^4 \over Q_1^2} + r_H^2 \,,
}
so the horizon radius $r_H$ remains an explicit parameter.  

Perhaps counterintuitively, the ratio of chemical potentials $\mu_R \equiv \mu_1/\mu_2$ at extremality decreases as $Q_1/Q_2$ (or $Q_1/r_H$) increases,
\eqn{RatioChemPots}{
\mu_R \equiv {\mu_1 \over \mu_2}
\;= \;\sqrt{{1 \over 8} \left(Q_1 \over Q_2\right)^2 + 1} - {1 \over \sqrt{8}} {Q_1 \over Q_2}
\; =\; {1 \over \sqrt{1 + {1 \over 2} \left( Q_1 \over r_H\right)^2}}
\,,
}
while the ratio of charge densities
\eqn{}{
{\rho_1\over \rho_2} = {Q_1\over \sqrt{2}Q_2}\,,
}
more expectedly moves in the other direction.
Furthermore, we see the expression \eno{RatioChemPots} is bounded above at $\mu_R = 1$; no extremal black holes in this class exist with larger ratio.  The $Q_1/Q_2 \to 0$ and $Q_1/Q_2 \to \infty$ (or $Q_1/r_H \to 0$ and $Q_1/r_H \to \infty$) limits of the extremal solutions are related to the 1-charge and 2-charge black holes, as we describe momentarily.

The 3-charge black hole is the equal-charge special case of a 2+1-charge solution:
\eqn{}{
Q_1 = Q_2 \equiv Q.
}
which holds when
\eqn{}{
A_\mu = a_{\mu} \quad \to \quad \Phi_2 = \Phi_1 \,.
}
The scalar field thus vanishes for the 3-charge solution, which solves the simplfied effective action
\eqn{}{
e^{-1} {\cal L} =    R + {12 \over L^2} - 3 f_{\mu\nu} f^{\mu\nu}
- 2 \epsilon^{\mu\nu\rho\sigma\tau} f_{\mu\nu} f_{\rho\sigma} a_\tau \,.
}
The 3-charge black hole has an extremal limit as well, with $r_H^2 = Q^2/2$, corresponding to 
\eqn{}{
\mu_R = {1 \over \sqrt{2}} \,.
}
Unlike the 2-charge and 1-charge black holes we describe next, the 3-charge geometry behaves as just another member of the general class of 2+1-charge black holes.

\subsection{2-charge black hole}

The 2-charge black hole comes from setting $Q_1 =0$ in the general 2+1-charge solution \eno{TwoPlusOneSoln}; this removes the gauge field $a_\mu$ from the system.
  The solution is
\eqn{}{
A(r) &= \log {r \over L} + {1 \over 3} \log \left( 1 + {Q_2^2\over r^2} \right)\,, \quad
B(r) = - \log {r \over L}- {2 \over 3} \log \left( 1 + {Q_2^2\over r^2} \right)\,, \cr
h(r) &= 1- { (r_H^2 + Q_2^2)^2  \over (r^2 + Q_2^2)^2 }\,, \quad
\phi(r) =  \sqrt{2\over 3} \log \left( 1 + {Q_2^2\over r^2} \right) \,, \quad
\Phi_2(r) ={Q_2 \over2  L} \left(1-{r_H^2 + Q_2^2 \over r^2 + Q_2^2} \right) \,.
}
The thermodynamics is
\eqn{TwoThermo}{
T = { r_H  \over  \pi L^2 } \,, \quad
\mu_2 = {\sqrt{2} Q_2  \over L^2} \,, \quad
s = {r_H \over 4 G L^3} (r_H^2 + Q_2^2) \,, \quad
\rho_2 = {\sqrt{2} Q_2 s \over 2 \pi r_H} \,,
}
with $\mu_1 = \rho_1 = 0$.
There is an extremal solution for $r_H =0$; most interestingly for us, we
notice that in this limit the entropy vanishes,
\eqn{}{
s_{T = 0} = 0\,.
}
For the extremal solution the scalar diverges at the horizon, suggesting new light modes appear and the geometry is strictly IR incomplete; however the missing modes are expected to be associated with unbroken abelian gauge symmetry based on a D3-brane configuration in ten dimensions \cite{Kraus:1998hv,Freedman:1999gk} (where the scalar describes the squashing of the $S^5$) which are ${\cal O}(N)$ and so subleading.

\subsection{1-charge black hole}
\label{sec:OneCharge}

The 1-charge black hole, conversely, comes from setting $Q_2 =0$, which removes the gauge field $A_\mu$ from the problem.  The solution is
\eqn{}{
A(r) &= \log {r \over L} + {1 \over 6} \log \left( 1 + {Q_1^2\over r^2} \right)\,, \quad
B(r) = - \log {r \over L}- {1 \over 3} \log \left( 1 + {Q_1^2\over r^2} \right)\,, \cr
h(r) &= 1- {r_H^2(r_H^2 + Q_1^2) \over  r^2(r^2 + Q_1^2) }\,, \quad
\phi(r) =  -\sqrt{2\over 3} \log \left( 1 + {Q_1^2\over r^2} \right) \,, \quad
\Phi_1(r) ={Q_1 r_H \over 2 L \sqrt{r_H^2 + Q_1^2}} \left(1-{r_H^2 + Q_1^2 \over r^2 + Q_1^2} \right)  \,.
}
The temperature and remaining chemical potential become
\eqn{}{
T = {2 r_H^2 + Q_1^2   \over 2 \pi L^2 \sqrt{r_H^2 + Q_1^2}} \,, \quad
\mu_1 = {Q_1 r_H \over L^2 \sqrt{r_H^2 + Q_1^2}}\,,
}
with $\mu_2 = 0$, 
and the entropy and nonzero charge density are
\eqn{}{
s = {r_H^2 \over 4 G L^3} (r_H^2 + Q_1^2)^{1/2}  \,, \quad \rho_1 = {Q_1 s \over 2 \pi r_H} \,.
}
Note there is no longer any solution for an extremal black hole. Instead, the limit $r_H \to 0$ of the 1-charge black hole is a BPS configuration with no gauge fields and no horizon function, preserving four-dimensional Lorentz invariance with only the scalar running.  It is actually the so-called $n=2$ renormalization group flow solution studied in \cite{Freedman:1999gk}, corresponding to a zero temperature and zero chemical potential state on the Coulomb branch of ${\cal N}=4$ Super-Yang-Mills; our scalar is related to the one in \cite{Freedman:1999gk} by $\phi = - 2\mu$, the parameter $\ell$ of that solution is simply $Q_1$ and the variable $v$ used there is given by
\eqn{vDef}{
v = {r^2 \over r^2 + Q_1^2} \,.
}
The 1-charge black hole is also a Poincar\'e patch limit of superstars considered in \cite{Myers:2001aq}.  Again the scalar diverges and the geometry is IR incomplete, but the lift is known to be a disc of D3-branes  \cite{Freedman:1999gk} and the missing modes should again be ${\cal O}(N)$ in number.

\subsection{Relation of extremal geometries}
\label{ExtremalRelationSec}

Up to overall rescaling of coordinates, the extremal 2+1-charge black holes are a one-parameter family of solutions determined by $Q_1/r_H$ (or equivalently $Q_1/Q_2$ or $Q_2/r_H$).   It would seem natural that the extreme limits of the one-parameter family are the extremal 2-charge black hole on one side and the Coulomb branch solution on the other.  In fact, order of limits issues make these identifications somewhat subtle.

The Coulomb branch solution follows from taking $Q_2 \to 0$ strictly to reach the 1-charge black hole, without first requiring extremality, and only afterwards taking $r_H \to 0$. Thus the ratios of the parameters are
\eqn{CoulombBranchDefined}{
{Q_2 \over Q_1} \to 0 \,, \quad \quad {r_H \over Q_1 }\to 0 \,, \quad \quad {Q_2 \over r_H} \to 0 \quad \quad \quad \quad \hbox{(Coulomb branch)} \,.
}
On the other hand, if we first take the extremal limit and afterward take $Q_2 \to 0$ with  $Q_1$ held finite, then $Q_2$ and $r_H$ are linked by the extremality condition \eno{ExtremalRule}, and we have
\eqn{}{
{Q_2 \over Q_1} \to 0 \,, \quad \quad {r_H \over Q_1 }\to 0 \,, \quad \quad {Q_2 \over r_H} \to 1 \quad \quad \quad \quad ({\rm extremal \; 2+1}) \,.
}
The same ratios result from $Q_1 \to \infty$ with $r_H$ fixed when one has eliminated $Q_2$ for extremality using \eno{Q2ExtremalRule}.  The Coulomb branch solution also has $\mu_2 = 0$, while the limit of the 2+1 black holes has instead $\mu_1/\mu_2 \to 0$.

Thus the limits are not the same. As it happens, despite the difference in the limits, the metric, gauge field $\Phi_1$ and scalar of the 2+1-charge extremal black holes nonetheless approach the Coulomb branch solution in the $r_H \to 0$ limit.  The only field that does not is $\Phi_2$, which instead of approaching $\Phi_2 \to 0$ as one should have for the 1-charge black hole, instead achieves the nonzero value
\eqn{}{
\Phi_2 \to  {Q_1 \over 2L} \,.
}
Thus a general field in the $r_H \to 0$ limit of the 2+1-charge extremal black holes will not explore the Coulomb branch geometry.  A field that is unaware of $\Phi_2$, however, will find its dynamics approaching that of the Coulomb branch. We will see this come in handy later.

There is a similar story for the 2-charge extremal black hole.  The true 2-charge extremal case involves $Q_1 \to 0$ strictly, followed by $r_H \to 0$ with $Q_2$ finite. This implies
\eqn{Extremal2}{
{Q_1 \over Q_2} \to 0 \,, \quad \quad {r_H \over Q_2 }\to 0 \,, \quad \quad {Q_1 \over r_H} \to 0 \quad \quad \quad \quad ({\rm extremal \; 2}) \,.
}
Indeed we have in this case
\eqn{}{
{Q_1 Q_2^\beta \over r_H^\alpha} \to 0 \quad \quad \quad \quad ({\rm extremal \; 2}) \,,
}
for any $\alpha$ and $\beta$. In addition the chemical potentials are $\mu_1 = 0$, $\mu_2 \neq 0$.

On the other hand, if we first take the extremal limit of the 2+1-charge black holes and then take $Q_1 \to 0$ with $r_H$ finite, we find $Q_2 \to \infty$.  Although the relations \eno{Extremal2} all follow in this case, we also have 
\eqn{QProduct}{
{Q_1 Q_2 \over r_H^2} \to \sqrt{2} \quad \quad \quad \quad ({\rm extremal \; 2+1}) \,,
}
whereas in the true 2-charge black hole, $Q_1 Q_2 / r_H^2 = 0$.  In addition the chemical potentials approach $\mu_1/\mu_2 \to 1$.  Thus again the limits are not the same.

Examining the solutions, again only one field fails to agree in the two limits: this time it is $\Phi_1$, which is zero in the true 2-charge black hole, but in the limit described around \eno{QProduct} approaches
\eqn{}{
\Phi_1 \to  {Q_2 \over \sqrt{2} L} \,.
}
Any fermion insensitive to the $a_\mu$ gauge field would thus not notice the difference in limits; however, unlike the previous case, we will have no such fermions.

\subsection{Field theory duals}
\label{FieldTheorySec}

The dual field theory to the extremal charged black hole solutions is ${\cal N}=4$ Super-Yang-Mills with gauge group $SU(N)$ at large number of colors and large 't Hooft coupling, at zero temperature and nonzero chemical potentials $\mu_1$ and $\mu_2$.  

The field content for ${\cal N}=4$ SYM is a gauge field, adjoint Majorana fermions  $\lambda_a$, $a = 1 \ldots 4$ in the ${\bf 4}$ of $SO(6)$, and adjoint scalars $X$ in the ${\bf 6}$ of $SO(6)$.  The charges of the dual gauginos and dual scalars $Z_j \equiv X_{2j - 1} + i X_{2j}$, $j = 1,2,3$ under the Cartan subgroup $U(1)_a \times U(1)_b \times U(1)_c$ are
\begin{equation}\begin{tabular}{|c|c|c|c|c|c|c|c|} \hline
 & $\lambda_1$ & $\lambda_2$   & $\lambda_3$ & $\lambda_4$ & $Z_1$ & $Z_2$ & $Z_3$ \\ \hline
$q_a$  & ${1 \over 2}$ & ${1 \over 2}$   & $- {1 \over 2}$ & $- {1\over 2}$ & $1$ & $0$ & $0$ \\
$q_b$  & ${1 \over 2}$ & $-{1 \over 2}$   & $ {1 \over 2}$ & $- {1\over 2}$ & $0$ & $1$ & $0$ \\
$q_c$  & ${1 \over 2}$ & $-{1 \over 2}$   & $- {1 \over 2}$ & ${1\over 2}$ & $0$ & $0$ & $1$ \\ \hline
\end{tabular}
\end{equation}
The charges $q_1 \equiv q_a$ and $q_2 \equiv q_b + q_c$ associated to the chemical potentials $\mu_1$ and $\mu_2$ are then
\begin{equation}\begin{tabular}{|c|c|c|c|c|c|c|c|} \hline
 & $\lambda_1$ & $\lambda_2$   & $\lambda_3$ & $\lambda_4$ & $Z_1$ & $Z_2$ & $Z_3$ \\ \hline
$q_1$  & ${1 \over 2}$ & ${1 \over 2}$   & $- {1 \over 2}$ & $- {1\over 2}$ & $1$ & $0$ & $0$ \\
$q_2$  & $1$ & $-1$   & $0$ & $0$ & $0$ & $1$ & $1$ \\ \hline
$q_3$  & ${3 \over 2}$ & $-{1 \over 2}$   & $- {1 \over 2}$ & $- {1\over 2}$ & $1$ & $1$ & $1$ \\ \hline
\end{tabular}
\end{equation}
where we also included $q_3 \equiv q_1 + q_2$.

The nonzero scalar field $\phi(r)$ in the gravity background \eno{TwoPlusOneSoln} leads to a vacuum expectation value for the dual operator in the ${\bf 20'}$,
\eqn{}{
{\cal O}_{\bf 20'} \sim {\rm Tr}\, (-2 |Z_1|^2 + |Z_2|^2 + |Z_3|^2 ) \,,
}
implying expectation values for the individual scalars.
The expectation value for ${\cal O}_{\bf 20'}$ vanishes at the 3-charge black hole $\mu_R = 1/\sqrt{2}$, and has opposite signs on either side of this point.  

We will introduce the fermionic fields we study in this paper and the operators they are dual to, including their charges,  in section~\ref{DualOperatorSec}.

\section{Charged black holes from maximal gauged supergravity}
\label{GaugedSUGRASec}

In order to derive the fermion fluctuation equations in the black brane geometries, we first obtain the embedding for the Lagrangian \eno{Lag} in the full maximal gauged supergravity theory.  This section is somewhat technical and may be skipped by a reader anxious to get to the results.

\subsection{Scalar coset representative}

The scalar fields of $D=5$, ${\cal N}=8$ gauged supergravity \cite{Gunaydin:1985cu} parameterize an 
$E_{6(6)}/USp(8)$ coset manifold, and the $SO(6)$ gauge group is embedded in both the numerator and denominator groups.  Under $SO(6)$ the scalars transform in the representations ${\bf 20'} \oplus {\bf 10} \oplus {\bf \overline{10}} \oplus {\bf 1} \oplus {\bf 1}$.
 The $E_{6(6)}$ is conveniently parameterized in terms of an $SL(6, {\mathbb R}) \times SL(2, {\mathbb R})$ subgroup, with $SO(6)$ embedded in $SL(6, {\mathbb R})$ in the natural way.
 Elements of the ${\bf 20'}$ sit in the coset $SL(6, {\mathbb R})/SO(6)$, while the singlet complex dilaton sits in the $SL(2, {\mathbb R})/SO(2)$ and the ${\bf 10} \oplus {\bf \overline{10}}$ sit in the off-diagonal blocks. 

The scalar arising in the 2+1-charge black holes lives in the ${\bf 20'}$ and may be described 
in terms of the $SL(6, {\mathbb R})$ element
\eqn{Smatrix}{
S = \diag \{e^{\phi \over \sqrt{6}},e^{\phi \over \sqrt{6}}, e^{- \phi \over 2\sqrt{6}}, e^{- \phi \over 2\sqrt{6}}, e^{- \phi \over 2\sqrt{6}}, e^{- \phi \over 2\sqrt{6}}\}\,,
}
while the $SL(2, {\mathbb R})$ element is $S' = {\mathbb I}$ and the off-diagonal terms are zero; $\phi(x)$ is the dynamical scalar field. Decomposing the $SL(6, {\mathbb R})$ indices $I, J = 1 \ldots 6$ into $x,y = 1,2$ and $i, j = 3, \ldots 6$ and denoting the $SL(2, {\mathbb R})$ index $\alpha, \beta$, we then have
\eqn{}{
S_x^{\;y} = e^{\phi \over \sqrt{6}} \delta_x^y \,, \quad \quad
S_i^{\;j} = e^{-{\phi \over2 \sqrt{6}}}\delta_i^j \,, \quad \quad
S_i^{\;x} = S_x^{\;i} =0 \,, \quad \quad (S')_\alpha^{\;\;\beta} = \delta_\alpha^\beta\,.
}
The $E_{6(6)}$ elements $U$ are defined as\footnote{By using the inverse of $S$ in the first term we correct a typo in equation (5.34) of \cite{Gunaydin:1985cu}.}
\eqn{}{
U^{IJ}_{\;\;\;\;KL} = 2 (S^{-1})_{[K}^{\;\;\;[I} (S^{-1})_{L]}^{\;\;\;J]} \,, \quad \quad
U^{IJK\alpha} = U_{K\alpha IJ} =0 \,, \quad \quad
U_{I\alpha}^{\;\;\;\;J\beta} = S_I^{\;\;J} (S')_\alpha^{\;\;\beta} \,,
}
and the scalar coset representatives are then
\eqn{CosetFormula}{
V^{IJab} &= {1 \over 8} \left[ (\Gamma_{KL})^{ab} U^{IJ}_{\;\;\;\;KL} + 2 (\Gamma_{K\beta})^{ab} U^{IJ\,K\beta} \right] \,, \cr
V_{I\alpha}^{\;\;\;ab} &= {1 \over 4 \sqrt{2}} \left[ (\Gamma_{KL})^{ab} U_{I\alpha  \, KL} + 2 (\Gamma_{K\beta})^{ab} U_{I\alpha}^{\;\;\;\;K\beta} \right] \,.
}
Here $a,b = 1 \ldots 8$ are $USp(8)$ indices 
 raised and lowered with the symplectic metric $\Omega_{ab}$ and its inverse $\Omega^{ab}$:
\eqn{}{
A_a = \Omega_{ab} A^b\,, \quad \quad A^a = \Omega^{ab} A_b \,,
}
where $\Omega^{-1} = \Omega^T = - \Omega$.\footnote{A counterintuitive fact is that $\Omega^{ab}$ is not what you get when you raise both indices of $\Omega_{ab}$ according to this rule; you get $\Omega^{ac} \Omega^{bd} \Omega_{cd} = \Omega^{ba}$.}
The $\Gamma_X$, $X = 0, 1, \ldots 6$ are hermitian, antisymmetric and pure imaginary $8 \times 8$ Euclidean $\Gamma$-matrices obeying 
\eqn{Clifford}{
\{ \Gamma_X, \Gamma_Y \} = 2 \delta_{XY} \,,
}
and related to each other by
\eqn{}{
\Gamma_0 = i  \Gamma_1 \cdots \Gamma_6  \,.
}
The $\Gamma$-matrices act on the $USp(8)$ space as Clebsch-Gordan coefficients.
Following \cite{Gunaydin:1985cu}, we separate out $\Gamma_I$, $I =1 \ldots 6$ from $\Gamma_0$, and define
\eqn{}{
\Gamma_{IJ} \equiv {1 \over 2} (\Gamma_I \Gamma_J - \Gamma_J \Gamma_I) \,, \quad \quad
\Gamma_{I \alpha} \equiv (\Gamma_I, i \Gamma_I \Gamma_0) \,,
}
where $\alpha = 1, 2$.  It is straightforward to show that $\Gamma_{IJ}$ and $\Gamma_{I \alpha}$ are also antisymmetric as $8 \times 8$ matrices. 
The $USp(8)$ metric  $\Omega_{ab}$ may be written as 
\eqn{OmegaMetric}{
\Omega_{ab} = - \Omega^{ab} = i (\Gamma_0)^{ab} \,.
}
The coset representatives \eno{CosetFormula} then become
\eqn{CosetReps}{
V^{xyab} &= {1 \over 4} e^{-2 \phi \over \sqrt{6}} (\Gamma_{xy})^{ab} \,, \quad \quad
V^{ijab} = {1 \over 4} e^{\phi \over \sqrt{6}} (\Gamma_{ij})^{ab} \,, \quad \quad
V^{ixab}= {1 \over 4} e^{ -\phi \over2 \sqrt{6}} (\Gamma_{ix})^{ab} \,, \cr
V_{x\alpha}^{\;\;\;ab} &=   {1 \over 2 \sqrt{2}} e^{\phi \over \sqrt{6}} (\Gamma_{x \alpha})^{ab} \,, \quad \quad
V_{i \alpha}^{\;\;\;ab} =  {1 \over 2 \sqrt{2}} e^{- \phi \over 2\sqrt{6}} (\Gamma_{i \alpha})^{ab}  \,,
}
and the inverses are
\eqn{}{
\tilde{V}_{abxy} &= {1 \over 4} e^{2 \phi \over \sqrt{6}} (\Gamma_{xy})^{ab} \,, \quad \quad
\tilde{V}_{abij} = {1 \over 4} e^{-\phi \over \sqrt{6}} (\Gamma_{ij})^{ab} \,, \quad \quad
\tilde{V}_{abix}={1 \over 4} e^{ \phi \over 2\sqrt{6}} (\Gamma_{ix})^{ab} \,, \cr
\tilde{V}_{ab}^{\;\;\;x\alpha} &=  - {1 \over 2 \sqrt{2}} e^{-\phi \over \sqrt{6}} (\Gamma_{x \alpha})^{ab} \,, \quad \quad
\tilde{V}_{ab}^{\;\;\;i \alpha} = - {1 \over 2 \sqrt{2}} e^{ \phi \over 2\sqrt{6}} (\Gamma_{i \alpha})^{ab}  \,.
}

\subsection{Composite connection and scalar kinetic terms}

The covariant derivative of $D=5$, ${\cal N}=8$ gauged supergravity acts on both $USp(8)$ indices with a composite connection $Q_{\mu a}^{\;\;\;\;b}$, and on $SO(6)$ indices with the fundamental gauge fields $A_{\mu IJ} \equiv A_{\mu [IJ]}$:
\eqn{}{
D_\mu X_{aI} &= \partial_\mu X_{aI} +Q_{\mu a}^{\;\;\;\;b} X_{bI} - g A_{\mu I}^{\;\;\;\;J} X_{aJ} \,, \cr 
D_\mu X^{aI} &= \partial_\mu X^{aI} - Q_{\mu b}^{\;\;\;\;a} X^{bI} - g A^{\;\;\;I}_{\mu \;\;J} X^{aJ} \,.
}
The covariant derivative of the coset representative defines the quantity $P_\mu^{\;\;abcd}$,
\eqn{PDef}{
P_\mu^{\;\;abcd} \equiv \tilde{V}^{abAB} D_\mu V_{AB}^{\;\;\;cd} \,,
}
where $AB$ is a formal index pair in the 27 of $E_{6(6)}$, whose contraction is defined as
\eqn{FormalSum}{
 \tilde{V}^{abAB} D_\mu V_{ABab}  \equiv  \tilde{V}^{ab}_{\;\;\;IJ} D_\mu V^{IJ}_{\;\;\;\;ab} +  \tilde{V}^{abI \alpha} D_\mu V_{I \alpha ab} \,,
}
and where the composite connection is defined so that $P_\mu^{\;\;abcd}$ is totally antisymmetric and symplectic traceless, denoted $P_\mu^{\;\;abcd} = P_\mu^{\;\;[abcd]|}$; this leads to
\eqn{}{
Q_{\mu a}^{\;\;\;\;\;b} = - {1 \over 3} \left[ \tilde{V}^{bcAB} \partial_\mu V_{ABac}  + g A_{\mu\, IL} \delta^{JL} (2 V_{\;\;\;\;\;ac}^{IK}\, \tilde{V}^{bc}_{\;\;\;JK} - V_{J\alpha ac} \, \tilde{V}^{bcI\alpha} )\right] \,.
}
In our background the composite connection is
\eqn{QResult}{
Q_{\mu a}^{\;\;\;\;\;b}  = - {g\over 2} a_\mu (\Gamma_{12})^{ab} - {g \over 2} A_\mu(\Gamma_{34} + \Gamma_{56})^{ab} \,.
}
The scalar kinetic terms in the gauged supergravity theory are 
\eqn{}{
{\cal L}_{\rm scalar} = {1 \over 24} P_{\mu abcd}\, P^{\mu abcd} \,,
}
and $P_\mu^{\;\;abcd}$ evaluates to
\eqn{}{
P_\mu^{\;\;abcd} &={1 \over 8 \sqrt{6}} \partial_\mu \phi \Bigg(
- (\Gamma_{xy})^{ab}(\Gamma_{xy})^{cd}
+ {1 \over 2} (\Gamma_{ij})^{ab}(\Gamma_{ij})^{cd}
- {1 \over 2} (\Gamma_{ix})^{ab}(\Gamma_{ix})^{cd}\cr
& + (\Gamma_{x})^{ab}(\Gamma_{x})^{cd}
-  (\Gamma_0 \Gamma_{x})^{ab}(\Gamma_0 \Gamma_{x})^{cd}
- {1 \over 2} (\Gamma_{i})^{ab}(\Gamma_{i})^{cd}
+ {1 \over 2} (\Gamma_0 \Gamma_{i})^{ab}(\Gamma_0 \Gamma_{i})^{cd}
\Bigg) \,,
}
leading to the kinetic term for the neutral scalar $\phi$,
\eqn{}{
{\cal L}_{\rm scalar} = {1 \over 8} (\partial \phi)^2 \,.
}
As we will see, this unconventional normalization will become canonical when we rescale the gauged supergravity Lagrangian to match \eno{Lag}.

\subsection{Scalar potential}

The gravity/scalar sector is then given by the Lagrangian
\eqn{}{
e^{-1} {\cal L}_{\rm gravity + scalar} = -{1 \over 4} R + {1 \over 8} (\partial \phi)^2 - V(\phi) \,.
}
The potential may be determined in terms of a few $USp(8)$ tensors that depend on the scalar; these tensors will also enter into the fermion couplings.
These start with the  tensor $T^a_{\;\;bcd}$,
\eqn{}{
T^a_{\;\;bcd} \equiv (2 V^{IKae}\, \tilde{V}_{beJK} - V_{J\alpha}^{\;\;\;\; ae} \, \tilde{V}_{be}^{\;\;\;\; I\alpha} ) \,  \tilde{V}_{cdIJ }\,,
}
which is evaluated to be (with one index lowered),
\eqn{}{
T_{abcd} = {3i \over 16} e^{-\phi\over \sqrt{6}} (\Gamma_0 \Gamma_{I i})^{ab} (\Gamma_{I i})^{cd}
+  {3 i\over 16} e^{2\phi\over \sqrt{6}} (\Gamma_0 \Gamma_{I x})^{ab} (\Gamma_{I x})^{cd} \,.
}
It is easy to see that this is symplectic traceless in the first pair of indices and in the second pair,
\eqn{}{
\Omega^{ab} T_{abcd} = \Omega^{cd} T_{abcd} = 0 \,,
}
as well as antisymmetric in the second pair but symmetric in the first,
\eqn{}{
T_{(ab)cd} = T_{ab[cd]} = T_{abcd} \,.
}
The tensor $T_{ab}$ is defined as
\eqn{}{
T_{ab} \equiv T^c_{\;\;abc} \,,
}
and evaluates to be
\eqn{RicciT}{
T_{ab} = \left( - {15 \over 4} e^{-\phi\over \sqrt{6}}  - {15 \over 8} e^{2\phi\over \sqrt{6}} \right) \delta_{ab}  \,,
}
and finally the tensor $A_{bcd}$ is defined as the antisymmetrization and trace-removal over the final three indices of $T_{abcd}$,
\eqn{}{
A_{abcd} \equiv T_{a[bcd]|} \,,
}
which can be shown with a little calculation to be
\eqn{WeylA}{
A_{abcd} = {1 \over 3} \left( T_{abcd} + T_{acdb} + T_{adbc} \right) + {1 \over 9} \left( T_{ab} \Omega_{cd} + T_{ac} \Omega_{db} + T_{ad} \Omega_{bc} \right) \,.
}
It is also straightforward to see that $T^{abcd}$, $T^{ab}$ and $A^{abcd}$ have the same values for their components as $T_{abcd}$, $T_{ab}$ and $A_{abcd}$.
The five-dimensional ${\cal N}=8$ gauged supergravity potential is then defined in all generality as
\eqn{ScalarPotential}{
V = {g^2 \over 96} \left( A_{abcd} A^{abcd} - {64 \over 225} T_{ab} T^{ab} \right) \,,
}
and is in our case
\eqn{Potential}{
V(\phi) = - {g^2 \over 2} e^{\phi \over \sqrt{6}} - {g^2 \over 4} e^{-2\phi\over \sqrt{6} } \,.
}
The superpotential $W(\phi)$ may be defined as  $W_{ab} \equiv W(\phi) \delta_{ab}$, where
\eqn{}{
W_{ab} \equiv {4 \over 15} T_{ab} \,,
}
becoming in our case 
\eqn{Superpotential}{
W(\phi) = - e^{- \phi \over \sqrt{6}} - {1 \over 2} e^{2\phi \over \sqrt{6}} \,.
}
The potential may be recovered from the superpotential by the usual gauged SUGRA formula
\eqn{}{
V(\phi) = {g^2 \over 8} \left( \partial W(\phi) \over \partial \phi \right)^2 - {g^2 \over 3} W(\phi)^2 \,.
}

\subsection{Gauge fields}

Turning to the gauge fields, 
the gauge kinetic term coming from ${\cal N}=8$ SUGRA is
\eqn{}{
e^{-1} {\cal L}_{\rm gauge \, kin} = - {1 \over 8} F_{\mu\nu ab}  F^{\mu \nu a b} \,,
}
where we have set the two-form fields to zero, and
\eqn{}{
F_{\mu\nu}^{\;\;\;\;ab} \equiv F_{\mu\nu IJ} V^{IJ ab} \,.
}
Using our expressions \eno{CosetReps} for the coset representatives, we obtain for the three Cartan gauge fields $A_{\mu 12}$, $A_{\mu 34}$, $A_{\mu 56}$, 
\eqn{}{
e^{-1} {\cal L}_{\rm gauge \, kin} 
&= - {1 \over 4} e^{-4\phi \over \sqrt{6}} F_{\mu\nu 12} F^{\mu\nu 12}  - {1 \over 4} 
e^{2\phi \over \sqrt{6}} F_{\mu\nu 34} F^{\mu\nu 34} - {1 \over 4} 
e^{2\phi \over \sqrt{6}} F_{\mu\nu 56} F^{\mu\nu 56} \,.
}
Let us define
\eqn{}{
a_ \mu \equiv A_{\mu 12} \,, \quad \quad A_\mu \equiv A_{\mu 34} = A_{\mu 56} \,,
}
and then we have
\eqn{}{
e^{-1} {\cal L}_{\rm gauge \, kin}= 
- {1 \over 4} e^{-4\phi \over \sqrt{6}} f_{\mu\nu} f^{\mu\nu}  - {1 \over 2} 
e^{2\phi \over \sqrt{6}} F_{\mu\nu} F^{\mu\nu}  \,.
}
There is also a Chern-Simons term,
\eqn{}{
{\cal L}_{\rm CS} = - {1 \over 96} \epsilon^{\mu\nu\rho\sigma\tau} \epsilon^{IJKLMN} F_{IJ\mu\nu} F_{KL\rho\sigma} A_{MN\tau} \,,
}
where we have dropped ${\cal O}(F A^3)$ and ${\cal O}(A^5)$  terms which vanish for us.  This evaluates to
\eqn{}{
{\cal L}_{\rm CS} = - {1 \over 6} \epsilon^{\mu\nu\rho\sigma\tau} \left( 2 f_{\mu\nu} F_{\rho\sigma} A_\tau + F_{\mu\nu} F_{\rho\sigma} a_\tau \right)
= - {1 \over 2} \epsilon^{\mu\nu\rho\sigma\tau} f_{\mu\nu} F_{\rho\sigma} A_\tau  \,,
}
where in the second equality we integrated by parts and dropped a surface term.
The complete gravity/scalar/gauge Lagrangian is thus
\eqn{}{
e^{-1} {\cal L} &= -{1 \over 4} R + {1 \over 8} (\partial \phi)^2 + g^2 \left( {1 \over 2} e^{\phi \over \sqrt{6}} + {1 \over 4} e^{-2\phi\over \sqrt{6} } \right)
- {1 \over 4} e^{-4\phi \over \sqrt{6}} f_{\mu\nu} f^{\mu\nu}  - {1 \over 2} 
e^{2\phi \over \sqrt{6}} F_{\mu\nu} F^{\mu\nu}  
- {1 \over 2} \epsilon^{\mu\nu\rho\sigma\tau}  f_{\mu\nu} F_{\rho\sigma} A_\tau  \,,
}
If we identify the gauged supergravity coupling $g$ with the AdS radius $L$, 
\eqn{OtherFields}{
g \equiv {2 \over L} \,,
}
the Lagrangian precisely matches \eno{Lag} up to an overall factor $1/4$, which can be absorbed into a definition of the gravitational constant.

\section{Fermionic fluctuation equations}
\label{FermionSec}

We now study the Lagrangian for spin-1/2 fields in ${\cal N}=8$ gauged supergravity, and extract the Dirac equations for fermionic fluctuations in the backgrounds of the black holes discussed in section~\ref{BlackHolesSec}.  A reader wishing to skip the details of gauged supergravity may proceed to the general Dirac equation~\eno{DiracMostlyMinus}.

\subsection{Spin-1/2 Dirac equations}

The spin-1/2 fields are the 48 $\chi^{abc}$, which are both antisymmetric and symplectic traceless, 
$\chi^{[abc]|} = 0$.   The 48 fermions are complex, but not independent; they are related by the symplectic Majorana condition:
\eqn{}{
\chi^{abc} = C (\bar\chi^{abc})^T \,.
}
where $C$ is the conjugation matrix defined via $(\gamma^\mu)^T = C \gamma^\mu C^{-1}$.
The barred version of the fermi field  is defined via conjugation as
\eqn{}{
\bar\chi^{abc} = (\chi_{abc})^\dagger \gamma^0 \,.
}
Note this definition relates a conjugate fermion with indices up to the original fermion with indices down.  Thus $\chi^{abc}$ and $\chi_{abc}$, which are {\em a priori} independent fields thanks to raising and lowering by the symplectic matrix $\Omega_{ab}$ (for example, $\chi^{567} = \chi_{123} \neq \chi_{567}$) are related as one another's conjugates by the symplectic Majorana condition.
Hence for the 48 spin-1/2 fields, only 24 are independent; the remaining are related to the conjugates of the first 24.  Completely analogous relations hold for the 8 gravitini $\psi_\mu^a$.

 The kinetic term for the fermi fields is
\eqn{}{
e^{-1} {\cal L}_{\rm kin} ={i \over 12} \bar\chi^{abc} \gamma^\mu D_\mu \chi_{abc}\,,
}
where the covariant derivative acting on the $\chi^{abc}$ is
\eqn{}{
  D_\mu \chi_{abc} = \nabla_\mu \chi_{abc} + Q_{\mu a}^{\;\;\;\;d}\,  \chi_{dbc}
+ Q_{\mu b}^{\;\;\;\;d}\,  \chi_{adc}+ Q_{\mu c}^{\;\;\;\;d}\,  \chi_{abd} \,,
}
where $\nabla_\mu$ contains the spin connection and $Q_{\mu a}^{\;\;\;\;\;b}$ is the composite connection  \eno{QResult}; this provides the gauge couplings for the fermions.
The spin-1/2 fields also have mass terms of the form,
\eqn{}{
e^{-1} {\cal L}_{\rm mass} = {i g \over 2} \bar\chi^{abc} \left( {1 \over 2} A_{bcde} - {1 \over 45} \Omega_{bd} T_{ce} \right) \chi_a^{\;\;de} \,, 
}
where $T_{ab}$ \eno{RicciT} and $A_{abcd}$ \eno{WeylA} are tensors on the scalar space defined previously, as well as Pauli terms 
\eqn{}{
e^{-1} {\cal L}_{\rm Pauli} &= {i \over 8} F_{\mu\nu}^{\;\;\;\;ab}  \bar\chi_{acd} \gamma^{\mu\nu} \chi_b^{\;\;cd} \cr
&= {i \over 16} \left( e^{- 2\phi \over \sqrt{6}} f_{\mu\nu} (\Gamma_{12})^{ab}
+  e^{\phi\over \sqrt{6}} F_{\mu\nu} (\Gamma_{34} + \Gamma_{56})^{ab}
\right)\bar\chi_{acd} \gamma^{\mu\nu} \chi_b^{\;\;cd}  \,.
}
We are interested in spin-1/2 fields that do not couple to the gravitini; we will discuss how to assure this momentarily.  For such fields the
quadratic Lagrangian in a 2+1-charge black hole background  takes the form
\eqn{}{
e^{-1} {\cal L} = {1 \over 2} \vec{\bar\chi} \Big[
i \gamma^\mu \nabla_\mu {\bf K} - {\bf M} + \gamma^\mu a_\mu {\bf Q}_1 + \gamma^\mu A_\mu {\bf Q}_2  + i f_{\mu\nu} \gamma^{\mu\nu} {\bf P}_1 + i F_{\mu\nu} \gamma^{\mu\nu} {\bf P}_2
\Big] \vec\chi \,,
}
where the vector  on $\vec\chi$ stands for the 48 fields encoded in the $abc$ indices.
One may diagonalize and normalize the kinetic term ${\bf K}$ to the identity.  The eigenvalues of ${\bf M}$, ${\bf Q}_i$ and ${\bf P}_i$ may then be found.  The equation for a fermion eigenvector is then
\eqn{DiracMostlyMinus}{
\left[ i \gamma^\mu \nabla_\mu -  g \left( m_1 e^{-\phi \over \sqrt{6}} + m_2 e^{2\phi \over \sqrt{6}} \right) +g q_1 \gamma^\mu a_\mu +gq_2 \gamma^\mu A_\mu + i p_1 e^{-2\phi\over \sqrt{6}} f_{\mu\nu} \gamma^{\mu\nu} + i p_2 e^{\phi\over \sqrt{6}} F_{\mu\nu} \gamma^{\mu\nu} \right] \chi = 0 \,. 
}
Here we have extracted a factor of $g$ from the eigenvalues of ${\bf Q}_i$ and exponentials depending on the scalar from the eigenvalues of ${\bf P}_i$, as well as writing the scalar-dependence of the ${\bf M}$ eigenvalues,
\eqn{}{
m = g \left( m_1 e^{-\phi \over \sqrt{6}} + m_2 e^{2\phi \over \sqrt{6}} \right) \,.
}
The fermion is then characterized by the six rational numbers $(m_i, q_i, p_i)$ .

It is straightforward to show that if a field $\chi_{abc}$ obeys \eno{DiracMostlyMinus} with couplings $(m_i, q_i, p_i)$, then the conjugate field/Majorana partner $ C (\bar\chi^{abc})^T = \chi^{abc}$ obeys \eno{DiracMostlyMinus} with couplings $(-m_i, -q_i, p_i)$; the eigenvectors thus come in pairs.
A negative value of $m$ corresponds to a fermi field whose field theory dual has opposite chirality.  
In practice, one may work  only with positive masses, by changing the Clifford basis of the negative mass modes, flipping the sign of all $\gamma$-matrices.  This effectively changes the sign of $m$ and $p_i$.  Thus instead of $(-m_i, -q_i, p_i)$ one may solve the equation for the conjugate/partner field with $(m_i, -q_i, -p_i)$, bearing in mind that the chirality of the dual field is opposite.

\subsection{Dual operators and ``maximal" spin-1/2 eigenvectors}
\label{DualOperatorSec}

The 48 spin-1/2 fields transform in the ${\bf 20} + {\bf \overline{20}}+ {\bf 4} + {\bf \overline{4}}$ representations of $SO(6)$.
The fields in the ${\bf 20}$ are dual to ${\cal N}=4$ Super-Yang-Mills operators of the form,
\eqn{}{
\chi^{abc} \sim {\rm Tr}\, (\lambda X) \,.
}
where $\lambda$ is the gaugino in the ${\bf 4}$ and $X$ is the adjoint scalar in the ${\bf 6}$.  There are twelve such fields with the mass values $(m_1, m_2) = (1/2, -1/4)$, four with mass values $(-1/2, 3/4)$ and four with mass values $(-1/6, 5/12)$; for all these cases the mass with vanishing scalar becomes
\eqn{}{
m_{\bf 20}(\phi = 0) = {g \over 4} = {1 \over 2L} \,.
}
Meanwhile the spin-1/2 fields in the ${\bf 4}$ are dual to the operators
\eqn{}{
\chi \sim {\rm Tr}\, (F_+ \lambda) \,,
}
where $F_+$ is the self-dual part of the $SU(N)$ field strength.  These fields have mass values $(m_1, m_2) = (1/2, 1/4)$; this mass as a function of the scalar fields is proportional to the superpotential and for vanishing scalar becomes
\eqn{}{
m_{\bf 4}(\phi =0 ) = {3g \over 4} = {3 \over 2L} \,.
}
We would like to study spin-1/2 fields that do not mix in the quadratic Lagrangian with the 8 spin-3/2 gravitino fields $\psi_\mu^a$, which transform in the ${\bf 4} + {\bf \overline{4}}$.  A number of such fields can be determined  by exploiting the conservation of the preserved U(1) charges.  
Since the metric and the scalar $\phi$ are neutral, the quadratic fermion action can only couple fermi fields with equal and opposite charges under 
any preserved U(1)s.  So if we identify a field $\chi^{abc}$ that has a charge possessed by no $\psi^a_\mu$, we can be assured that $\chi^{abc}$ does not mix with the gravitino in the quadratic action.

We can find the most decoupled spin-1/2 fields by considering a more general background: the $U(1)_a \times U(1)_b \times U(1)_c$ solutions with $A_{\mu12}$, $A_{\mu34}$ and $A_{\mu56}$ all nonzero and distinct.  This background has two (neutral) scalar fields in general, but we will not need to consider the details of the solutions.  In these backgrounds, each spin-1/2 field has three $U(1)$ charges $q_a, q_b, q_c$ and the above argument holds for all three.  (In our background, $q_1 = q_a$ and $q_2 = q_b + q_c$.\footnote{Note this is different from the rule for the black hole's charge, which is $Q_2 = Q_b = Q_C$.})

 The ${\bf 4}$ decomposes into  $U(1)_a \times U(1)_b \times U(1)_c$ charge vectors $({1 \over 2}, {1 \over 2}, {1 \over 2}$), $({1 \over 2}, -{1 \over 2}, -{1 \over 2}$), $(-{1 \over 2}, {1 \over 2}, -{1 \over 2}$), and $(-{1 \over 2}, -{1 \over 2}, {1 \over 2}$), while the ${\bf \overline{4}}$ is the negative of these; thus all three U(1) charges for the gravitini have magnitude $1/2$.  The ${\bf 20}$, meanwhile, contains two copies of the weight vectors of the ${\bf 4}$, and twelve distinct weight vectors where one of the three charges has magnitude $3/2$.  Thus of the 48 spin-1/2 fields in 24 symplectic Majorana pairs, there are 12 pairs we can be assured do not mix with the gravitini due to having one ``maximal" $U(1)$ charge of magnitude $3/2$. 

The charges of the dual gauginos $\lambda_a$ and dual scalars $Z_j $ are given in the table in section~\ref{FieldTheorySec}.
A moment's examination reveals that the way to get  an operator with one charge of magnitude $3/2$ is to make sure the nonzero charge of the scalar has the same sign as the corresponding charge in the gaugino.

In the following table we list the 12 ``maximal" modes $\chi^{q_a q_b q_c}$ with positive $q_3$.  All the maximal-charge fields have one of two sets of mass values.  Some modes are right-handed; this is manifested when the eigenvalue of the mass operator comes out negative.  For such fields we use the opposite-sign $\gamma$ basis as previously discussed, flipping the sign of the mass and Pauli couplings; these have already been flipped in the table.  Such fields have right-handed duals and are indicated by $\bar\chi$ instead of $\chi$ in the first column.

\begin{equation}
\begin{tabular}{|c|c|c|c|c|c|c|c|} \hline
$\chi^{q_a q_b q_c}$ &Dual operator &$m_1$& $m_2$ & $q_1$& $q_2$  & $p_1$ & $p_2$\\ \hline
$\chi^{({3 \over 2}, {1 \over 2}, {1 \over 2})}$ &$\lambda_1 Z_1$ &$-{1 \over 2}$& ${3 \over 4}$ & ${3 \over 2}$& $1$ & $- {1 \over 4}$& ${1 \over 2}$\\
$\chi^{({3 \over 2}, -{1 \over 2}, -{1 \over 2})}$ & $\lambda_2 Z_1$&$-{1 \over 2}$& ${3 \over 4}$  & ${3 \over 2}$& $-1$ & $- {1 \over 4} $& $-{1 \over 2}$\\
$\bar\chi^{({3 \over 2}, -{1 \over 2}, {1 \over 2})}$  , $\bar\chi^{({3 \over 2}, {1 \over 2}, -{1 \over 2})}$ & $\overline\lambda_3 Z_1$, $\overline\lambda_4 Z_1$&$-{1 \over 2}$& ${3 \over 4}$  & ${3 \over 2}$& $0$ & $-{1 \over 4} $&$0$\\ \hline
$\chi^{({1 \over 2}, {3 \over 2}, {1 \over 2})}$, $\chi^{({1 \over 2}, { 1\over 2}, {3 \over 2})}$ & $\lambda_1 Z_2$, $\lambda_1 Z_3$&${1 \over 2}$& $-{1 \over 4}$  & ${1 \over 2}$& $2$ & ${1 \over 4} $& $0$\\
$\bar\chi^{(-{1 \over 2},  {3 \over 2}, {1 \over 2})}$, $\bar\chi^{(-{1 \over 2},  {1 \over 2}, {3 \over 2})}$ & $\overline\lambda_2 Z_2$, $\overline\lambda_2 Z_3$&${1 \over 2}$& $-{1 \over 4}$ & $-{1 \over 2}$& $2$ & $-{1 \over 4}$& $0$ \\ 
$\chi^{(-{1 \over 2},  {3 \over 2}, -{1 \over 2})}$, $\chi^{(-{1 \over 2}, -{1 \over 2},  {3 \over 2})}$& $\lambda_3 Z_2$, $\lambda_4 Z_3$&${1 \over 2}$& $-{1 \over 4}$& $-{1 \over 2}$& $1$ & $- {1 \over 4}$& $- {1 \over 2}$\\
$\bar\chi^{({1 \over 2}, -{1 \over 2},  {3 \over 2})}$, $\bar\chi^{({1 \over 2},  {3 \over 2}, -{1 \over 2})}$& $\overline\lambda_3 Z_3$, $\overline\lambda_4 Z_2$& ${1 \over 2}$& $-{1 \over 4}$ & ${1 \over 2}$& $1$ & ${1 \over 4}$& $-{1 \over 2}$ \\ \hline
\end{tabular}
\end{equation}
The fields $\chi^{({3 \over 2}, {1 \over 2}, {1 \over 2})}$, $\chi^{({1 \over 2},  {3 \over 2}, {1 \over 2})}$ and $\chi^{({1 \over 2},  {1 \over 2}, {3 \over 2})}$ have maximal $q_3 = 5/2$, and thus in the background of the 3-charge black hole are the modes studied in \cite{DeWolfe:2011aa}.

The conjugates/symplectic Majorana partners of the modes in the table have the opposite chirality, and the signs of the $q_i$ and $p_i$ flipped.

\subsection{``Overlapping" spin-1/2 eigenvectors}

The 24 remaining spin-1/2 fields have charges $(q_a q_b q_c) = (\pm {1 \over 2}, \pm {1 \over 2}, \pm {1\over 2})$, each of these 8 charge assignments occurring with multiplicity 3; these modes overlap with the gravitini, which have the same 8 charge vectors.  Although group theory does not rule out any of these fields from mixing with the gravitini, we can see explicitly that some do not.

There are three types of $\chi^{abc}/\psi_\mu^a$ couplings 
 in the gauged supergravity Lagrangian, a direct coupling mediated by the $A_{abcd}$ tensor, a coupling with a scalar derivative $\partial_\mu \phi$ coming from $P_{\mu abcd}$, and a Pauli-type term, which in our 2+1-charge background splits into two distinct couplings, one for each gauge field:
\eqn{GravitinoCouplings}{
e^{-1} {\cal L}_{3/2 + 1/2} = {i g \over 6 \sqrt{2}} A_{dabc} \bar\chi^{abc}\gamma^\mu \psi_\mu^d + {i \over 3 \sqrt{2}} P_{\nu abcd} \bar\psi_\mu^a \gamma^\nu \gamma^\mu \chi^{bcd} 
+ {i \over 4\sqrt{2}} F_{\mu\nu}^{\;\;\;ab} \bar\psi_\rho^c \gamma^{\mu\nu} \gamma^\rho \chi_{abc} \,.
}
We may diagonalize the mass matrix for each three-dimensional charge subsector of spin-1/2 fields. We find three distinct masses in each sector: one with mass values $(m_1, m_2) = (1/2, 1/4)$, one with $(1/2, -1/4)$ and one with $(-1/6, 5/12)$. Plugging the mass eigenvectors into the gravitino couplings \eno{GravitinoCouplings}, we find the $(1/2, 1/4)$ eigenvectors never couple to the gravitini; these are the higher-mass modes dual to Tr $F_+ \lambda$.

The $(1/2, -1/4)$ eigenvectors in this sector have zero for most gravitino couplings, but two out of four of them notice the $A_\mu$ Pauli-type coupling.  The $(-1/6, 5/12)$ eigenvectors, on the other hand, couple to the gravitino in almost every possible way; the only gravitino couplings they don't have are the two $A_\mu$ Pauli couplings just mentioned. 

In the table, the last four columns indicate whether the fermion in question has a gravitino coupling of the sort indicated.   Thus there are six distinct decoupled Dirac equations from this sector; the ones with mass values $(1/2, -1/4)$, however, have the same Dirac equation as some of the ``maximal" modes.  Thus the only new Dirac equations we get from this sector are the four with mass $(1/2, 1/4)$ which are dual to Tr $(F_+ \lambda)$, three of which are distinct.

Again, the conjugate modes exist with opposite chirality and opposite signs for the $q_i$ and $p_i$.

\begin{equation}
\begin{tabular}{|c|c|c|c|c|c|c||c|c|c|c|} \hline
$\chi^{q_a q_b q_c}$ &$m_1$ & $m_2$ & $q_1$& $q_2$  & $p_1$ & $p_2$
& $A_{dabc} $ & $P_{\mu abcd}$& $f_{\mu\nu} \psi$ & $F_{\mu\nu} \psi $\\ \hline
$\chi_1^{({1 \over 2}, {1 \over 2}, {1 \over 2})}$  &${1 \over 2} $ & ${1 \over 4}$ & ${1 \over 2}$& $1$ & $- {1 \over 4}$& $-{1 \over 2}$ &---&---&---&---\\
$\bar\chi_2^{({1 \over 2}, {1 \over 2}, {1 \over 2})}$  &${1 \over 2} $ & $-{1 \over 4}$ & ${1 \over 2}$& $1$ & ${1 \over 4}$& $-{1 \over 2}$ &---&---&---&---\\
$\bar\chi_3^{({1 \over 2}, {1 \over 2}, {1 \over 2})}$   &$-{1 \over 6} $ & ${5 \over 12}$  & ${1 \over 2}$& $1$ & $- {5 \over 12}$& ${1 \over 6}$ &$\checkmark$&$\checkmark$&$\checkmark$&$\checkmark$\\ \hline
$\bar\chi_1^{(-{1 \over 2}, {1 \over 2}, {1 \over 2})}$  &${1 \over 2} $ & ${1 \over 4}$& $-{1 \over 2}$& $1$ & ${1 \over 4}$& $-{1 \over 2}$ &---&---&---&---\\
$\chi_2^{(-{1 \over 2}, {1 \over 2}, {1 \over 2})}$  &${1 \over 2} $ & $-{1 \over 4}$ & $-{1 \over 2}$& $1$ & $-{1 \over 4}$& $-{1 \over 2}$ &---&---&---&---\\
$\chi_3^{(-{1 \over 2}, {1 \over 2}, {1 \over 2})}$   &$-{1 \over 6} $ & ${5 \over 12}$ & $-{1 \over 2}$& $1$ & $ {5 \over 12}$& ${1 \over 6}$ &$\checkmark$&$\checkmark$&$\checkmark$&$\checkmark$\\ \hline
$\bar\chi_1^{({1 \over 2}, -{1 \over 2}, {1 \over 2})}$  &${1 \over 2} $ & ${1 \over 4}$ & ${1 \over 2}$& $0$ & $-{1 \over 4}$& $0$ &---&---&---&---\\
$\chi_2^{({1 \over 2}, -{1 \over 2}, {1 \over 2})}$  &${1 \over 2} $ & $-{1 \over 4}$  & ${1 \over 2}$& $0$ & ${1 \over 4}$& $0$ &---&---&---&$\checkmark$\\
$\chi_3^{({1 \over 2}, -{1 \over 2}, {1 \over 2})}$   &$-{1 \over 6} $ & ${5 \over 12}$ & ${1 \over 2}$& $0$ & $ -{5 \over 12}$& $0$ &$\checkmark$&$\checkmark$&$\checkmark$&---\\ \hline
$\bar\chi_1^{({1 \over 2}, {1 \over 2}, -{1 \over 2})}$  &${1 \over 2} $ & ${1 \over 4}$& ${1 \over 2}$& $0$ & $-{1 \over 4}$& $0$ &---&---&---&---\\
$\chi_2^{({1 \over 2}, {1 \over 2}, -{1 \over 2})}$  &${1 \over 2} $ & $-{1 \over 4}$  & ${1 \over 2}$& $0$ & ${1 \over 4}$& $0$ &---&---&---&$\checkmark$\\
$\chi_3^{({1 \over 2}, {1 \over 2}, -{1 \over 2})}$   &$-{1 \over 6} $ & ${5 \over 12}$& ${1 \over 2}$& $0$ & $ -{5 \over 12}$& $0$ &$\checkmark$&$\checkmark$&$\checkmark$&---\\ \hline
\end{tabular}
\end{equation}

\section{Solving Dirac equations, Fermi surfaces and oscillatory regions}
\label{FermiDiracSec}

In order to look for holographic Fermi surfaces in ${\cal N}=4$ Super-Yang-Mills theory, we must analyze the various Dirac equations we have obtained.  We do this in section~\ref{sec:Solving} for a slightly more general background than we later require.  Then we specialize to the $2+1$-charge extremal black hole backgrounds, characterized by $Q_1/r_H$, or equivalently, by $\mu_R \equiv \mu_1/\mu_2$, and present the near-horizon asymptotic analysis followed by a numerical study that allows us to trace out the main features of the holographic Fermi surfaces as a function of $\mu_R$.  In later sections we will consider more carefully the $1$-charge and $2$-charge limits.

\subsection{Solving the Dirac equation}
\label{sec:Solving}

We now analyze the  Dirac equation
\eqn{DiracEqn}{
\left( i \gamma^\mu \nabla_\mu -  m(\phi) +g q_1 \gamma^\mu a_\mu +gq_2 \gamma^\mu A_\mu + i p_1 e^{-2\phi\over \sqrt{6}} f_{\mu\nu} \gamma^{\mu\nu} + i p_2 e^{\phi \over \sqrt{6}} F_{\mu\nu} \gamma^{\mu\nu} \right) \chi = 0 \,,
}
in our class of backgrounds; we follow the basic conventions of \cite{Faulkner:2009wj}, though in opposite signature.  The covariant derivative contains the spin connection,
\eqn{}{
\nabla_\mu \equiv \partial_\mu - {1 \over 4} \omega_{\hat{a} \hat{b} \mu} \gamma^{\hat{a} \hat{b}} \,,
}
and we define
\eqn{ChiToPsi}{
\chi = e^{-2A} h^{-1/4} e^{-i\omega t + i k x} \Psi \,,
}
where for convenience we have placed the momentum $k$ in the $x$-direction, and the $e^{-2A} h^{-1/4}$ factor precisely cancels the effects of the spin connection.  Choosing a $\gamma$-matrix basis including
\eqn{}{
\gamma^{\hat{r}} =  \begin{pmatrix} i \sigma_3 & 0 \\  0 & i \sigma_3 \end{pmatrix} \,, \quad
\gamma^{\hat{t}} =  \begin{pmatrix} \sigma_1 & 0 \\  0 & \sigma_1\end{pmatrix} \,, \quad
\gamma^{\hat{i}} =  \begin{pmatrix} i \sigma_2  & 0 \\  0 & -i \sigma_2\end{pmatrix} \,,
}
we can define the projectors
\eqn{}{
\Pi_\alpha \equiv {1 \over 2} \left(1 - (-1)^{\alpha} i \gamma^{\hat{r}} \gamma^{\hat{t}} \gamma^{\hat{i}} \right) \,,\quad \quad
P_\pm \equiv {1 \over 2}\left ( 1 \pm i \gamma^{\hat{r}} \right) \,,
}
with $\alpha = 1,2$,
and  characterize the four components of $\Psi$ as
\eqn{}{
\Psi_{\alpha \pm} \equiv \Pi_\alpha P_\pm \Psi \,.
}
The Dirac equation decomposes into pairs relating $\Psi_{\alpha +}$ to $\Psi_{\alpha -}$ for each $\alpha$:
\eqn{}{
\left(\partial_r + {m e^B \over \sqrt{h} } \right)\Psi_{\alpha -} &= 
{ e^{B-A} \over \sqrt{h}} \left[ u(r) + (-1)^\alpha k -  v(r) \right] \Psi_{\alpha +} \,, \cr
\left(\partial_r - {m e^B \over \sqrt{h} } \right)\Psi_{\alpha +} &= 
{ e^{B-A} \over \sqrt{h}} \left[ -u(r) + (-1)^\alpha k - v(r)\right] \Psi_{\alpha -} \,,
}
where we have defined\footnote{One should not confuse $v(r)$ with the variable $v$ from \cite{Freedman:1999gk} defined in \eno{vDef} and used in section~\ref{RGFlowSec}.}
\eqn{}{
u(r) \equiv {1 \over \sqrt{h}} ( \omega + g q_1 \Phi_1 + g q_2 \Phi_2) \,, \quad \quad
v(r) \equiv 2 e^{-B} (p_1e^{-2\phi \over \sqrt{6}} \partial_r \Phi_1 + p_2 e^{\phi \over \sqrt{6}}  \partial_r \Phi_2) \,.
}
We can decouple these into second-order equations, obtaining
\eqn{SecondOrderDirac}{
 \Psi''_{\alpha \pm} - F_\pm  \Psi'_{\alpha \pm} 
+ 
\left[ \mp\partial_r \left(m e^{B} \over \sqrt{h} \right) - {m^2 e^{2B} \over h} 
+ {e^{2B-2A} \over h } \Big(u(r)^2 -(v(r) - (-1)^\alpha k)^2 \Big) \pm {m e^{B} \over \sqrt{h}} F_\pm \right]\Psi_{\alpha \pm}\,,
}
where
\eqn{}{
F_\pm \equiv \partial_r \log \left[{ e^{B-A} \over \sqrt{h}} \Big( v(r) - (-1)^\alpha k \pm u(r) \Big)  \right]\,.
}
One can see that \eno{SecondOrderDirac} is invariant under 
\eqn{SignFlips}{
q_i \to - q_i\,, \quad \quad p_i \to - p_i \,, \quad \quad \omega \to - \omega \,, \quad \quad k \to -k \,, 
}
and so the equation of the conjugate fermion has the same solutions with $(k, \omega) \to (-k, -\omega)$.

In the near-boundary limit $r \to \infty$, where all our geometries approach anti-de Sitter space, the Dirac equation is dominated by the mass term.    For $|mL| \neq 1/2$, we have the solutions
\eqn{BoundaryScalings}{
\Psi_{\alpha +} \sim A_\alpha(k) r^{mL} + B_\alpha(k) r^{-mL - 1} \,, \quad \quad
\Psi_{\alpha -} \sim C_\alpha(k) r^{mL-1} + D_\alpha(k) r^{-mL} \,.
}
The 
first order equations can be used to algebraically relate $B$ to $D$, and $A$ to $C$ \cite{Iqbal:2009fd},
\eqn{}{
C_\alpha = {L^2 (\omega + (-1)^\alpha k) \over 2mL-1} A_\alpha \,, \quad \quad
B_\alpha = {L^2 (\omega - (-1)^\alpha k) \over 2mL+1} D_\alpha \,.
}
For $m > 0$, $A$ is the source term, and $D$ the response; the case $m<0$  exchanges their roles.  Thus as we have mentioned previously, negative mass corresponds to opposite chirality in the dual.

For $mL = 1/2$, which is our primary case, we have instead
\eqn{}{
\Psi_{\alpha -} \sim C_\alpha(k) r^{-1/2} \log r + D_\alpha(k) r^{-1/2} \,,
}
with the relation between $B$ and $D$ unchanged but the relation between $A$ and $C$ now
\eqn{}{
C_\alpha = L^2 (\omega + (-1)^\alpha k) A_\alpha \,.
}
The retarded Green's function $G_R$ for the dual fermionic operator is defined in terms of the source $A$ and the response $D$ as 
\eqn{}{
D_\alpha = (G_R)_{\alpha \beta} A_\beta \,,
}
for a fluctuation with infalling boundary conditions imposed at the horizon.  Due to the decoupling of the $\alpha$ components in \eno{SecondOrderDirac}, the Green's function is diagonal for us, with $G_{22} (\omega, k) = G_{11} (\omega, -k)$.

\subsection{Near-horizon analysis: 2+1-charge case}

The extremal 2+1-charge black holes fit into a class of geometries with a double pole in the metric at the horizon, which has been studied in \cite{Faulkner:2009wj}.  As $r \to r_H$ we have the leading terms
\eqn{}{
g_{ii} \to k_0^2 \,, \quad \quad&
g_{tt} \to - \tau_0^2(r - r_H)^2 \,, \quad \quad
g_{rr} \to (L_2)^2 (r - r_H)^{-2}  \,, \cr
\Phi_i &\to \beta_i (r - r_H) \,, \quad \quad \phi \to \phi_0 \,,
}
where $k_0$, $\tau_0$, $L_2$, $\beta_i$ and $\phi_0$ are constants (we have eliminated $Q_2$ to impose extremality using \eno{Q2ExtremalRule}):
\eqn{}{
k_0 &= {2^{1/3} \over L} \left( r_H \over Q_1 \right)^{2/3} \sqrt{r_H^2 + Q_1^2} \,, \quad \quad
\tau_0 = {2^{1/3} \over L} \left( Q_1 \over r_H \right)^{1/3} \sqrt{4r_H^2 + Q_1^2 \over r_H^2 + Q_1^2} \,, \cr
L_2 &= {L \over 2^{2/3}} {Q_1^{1/3} r_H^{2/3} \over \sqrt{4r_H^2 + Q_1^2}} \,, \quad \quad
\phi_0 = \sqrt{2 \over 3} \log \left( 2 r_H^2 \over Q_1^2 \right) \,, \cr
\beta_1  &=  {2 r_H^2 \over L Q_1 \sqrt{r_H^2 + Q_1^2}} \,, \quad \quad 
\beta_2  = {Q_1 \sqrt{2 r_H^2 +  Q_1^2} \over 2 L r_H \sqrt{r_H^2 + Q_1^2}}\,.
}
Let us consider the Dirac equation in the near-horizon limit at $\omega =0$.
We have
\eqn{}{
u \to {g k_0 \over \tau_0} ( q_1 \beta_1 &+ q_2 \beta_2) \,, \quad \quad
 v \to {2 k_0 \over \tau_0 L_2} \left(p_1 e^{-2 \phi_0 \over \sqrt{6}} \beta_1 + p_2e^{ \phi_0\over\sqrt{6}} \beta_2\right) \,, \cr
 &m \to m(\phi_0) \equiv g \left( m_1 e^{- \phi_0 \over \sqrt{6}}+ m_2 e^{2 \phi_0 \over \sqrt{6}} \right)\,,
}
leading to
\eqn{}{
F_\pm \to - {1 \over r- r_H} \,.
}
The Dirac equation then simplifies to
\eqn{NearHorizonDirac}{
\partial_r^2 \Psi_{\alpha \pm} +{1 \over r- r_H}  \Psi_{\alpha \pm} 
- {\nu_k^2 \over (r - r_H)^2} \Psi_{\alpha \pm}\,,
}
where $\nu_k$ is
\eqn{nuk}{
\nu_k &= \sqrt{\left( m^2(\phi_0) + \left( \tilde{k}/ k_0\right)^2 \right) (L_2)^2 - g^2 \left( q_1 e_1 + q_2 e_2\right)^2}
}
with
\eqn{ktilde}{
\tilde{k} \equiv k - (-1)^\alpha {2 k_0 \over (L_2)^2} \left(p_1e^{-2 \phi_0 \over \sqrt{6}} e_1 + p_2e^{\phi_0 \over \sqrt{6}} e_2\right)  
\,.
}
and
where we have defined the $e_i$,
\eqn{}{
e_i \equiv {\beta_i L_2 \over \tau_0} \,. 
}
 taking the values
\eqn{}{
e_1 = {L r_H^3 \over Q_1 (4 r_H^2 + Q_1^2)} \,, \quad \quad
e_2 = {L Q_1 \sqrt{2 r_H^2 + Q_1^2} \over 4 (4 r_H^2 + Q_1^2)} \,, 
}
The solutions to \eno{NearHorizonDirac} are then simply
\eqn{HorizonBCs}{
\Psi \sim (r - r_H)^{\pm \nu_k} \,.
}
As described in \cite{Faulkner:2009wj}, these solutions possess a hidden near-horizon region approaching $AdS_2 \times \mathbb{R}^3$; appropriate boundary conditions are fixed by matching modes through this region. Defining the variables
\eqn{}{
r - r_H \equiv {\lambda L_2 \over \tau_0}{1 \over \zeta} \,, \quad \quad t \equiv {1 \over \lambda } \tau \,,
}
and taking the $\lambda \to 0$ limit with $\zeta$ and $\tau$ held fixed, we arrive at the metric
\eqn{}{
ds^2 = {(L_2)^2 \over \zeta^2 } \left( - d\tau^2 + d \zeta^2 \right) + k_0^2d\vec{x}^2 \,,
}
which is $AdS_2 \times \mathbb{R}^3$ with $AdS_2$ length $L_2$, while the gauge fields become
\eqn{}{
a_\mu \, dx^\mu = {e_1 \over \zeta} d\tau \,, \quad \quad 
A_\mu \, dx^\mu = {e_2 \over \zeta} d\tau \,,
}
identifying the $e_i$ as the coefficients of the gauge fields in the $AdS_2$ limit.    Imposing infalling boundary 
conditions in the $AdS_2$ geometry and matching out into the outer region requires picking the 
$\Psi \sim (r - r_H)^{+ \nu_k}$ boundary solution given above in \eno{HorizonBCs}.

\subsection{The oscillatory region, Fermi surfaces and excitations}

Consider the structure of the exponent $\nu_k$ \eno{nuk}.  The effect of the Pauli terms is to shift the origin of the 3-momentum $k$ to the quantity $\tilde{k}$ defined in \eno{ktilde}.  The quantity $\nu_k^2$ is then a sum of positive-definite contributions from the near-horizon mass and the shifted momentum $\tilde{k}$, and a negative-definite contribution from the effective near-horizon electric field coupling,
\eqn{}{
(q e)_{\rm eff} \equiv g q_1 e_1 + g q_2 e_2 \,.
}
When the electric coupling is sufficiently strong certain values of $k$ will produce an imaginary $\nu_k$.  This is called the {\em oscillatory region}, associated with an instability toward the pair production of charged particles in the $AdS_2$ region \cite{Pioline:2005pf}, and manifesting log-periodic behavior for the spinor excitations  \cite{Liu:2009dm, Faulkner:2009wj}.  The boundary of the oscillatory region occurs at $k_{\rm osc}$ satisfying
\eqn{kOsc}{
\nu_{k_{\rm osc}} = 0 \,.
}
From \eno{nuk}, it is easy to see that
\eqn{ktildeosc}{
\tilde{k}_{\rm osc}^2 = \left( k_0 \over L_2 \right)^2 \left( (qe)_{\rm eff}^2 - m^2 L_2^2 \right) \,,
}
and thus the oscillatory region may only exist when the effective electric coupling $(qe)_{\rm eff}^2$ has greater magnitude than the effective mass $m^2 L_2^2$.  It is thus significant that for the gauged supergravity modes we study, the mass depends on the scalar field and hence is a function of $\mu_R$; bottom-up constructions of fermion fluctuations with a constant mass miss this aspect of the spectrum.  (The effects of Pauli terms were considered in \cite{Edalati:2010ge}.)
As functions of $\mu_R$, these take the form
\eqn{GaugeAndMass}{
(qe)_{\rm eff}^2 = {(\sqrt{2} q_1 \mu_R^3 + q_2(1 - \mu_R^2))^2 \over 4 (1-\mu_R^2) (1 + \mu_R^2)^2} \,, \quad \quad
m^2 L_2^2 = { (m_1(1-\mu_R^2) + m_2 \mu_R^2)^2 \over (1 - \mu_R^4)} \,,}
while the proportionality factor is
\eqn{}{
\left( k_0 \over L_2\right)^2 = {4 r_H^2 (2 - \mu_R^2) (1 + \mu_R^2) \over L^4 \mu_R^2 (1 - \mu_R^2)}  = 2 \mu_2^2 (1 + \mu_R^2) \,,
}
where in the second equality we removed $r_H^2/L^4$ in favor of the chemical potentials $\mu_1$, $\mu_2$.
Meanwhile the shift between $k$ and $\tilde{k}$ \eno{ktilde} is opposite for the two components $\alpha = 1, 2$ and is determined by the $p_i$,
\eqn{kshift}{
k_{\rm shift} \equiv \tilde{k}- k &= -(-1)^\alpha {2 r_H (\sqrt{2} p_1 \mu_R+ p_2) \over L^2 \mu_R} \sqrt{2 - \mu_R^2 \over 1- \mu_R^2} \cr &=
(-1)^\alpha (2 p_1 \mu_1 + \sqrt{2} p_2 \mu_2) \,.
}
In plotting our results for $k_{\rm osc}$ and $k_F$, it is useful to normalize these dimensionful parameters by a quantity meaningful in the field theory.  The chemical potentials are a natural choice, and since $0 < \mu_1 < \mu_2$ for the extremal 2+1-charge black holes, the more convenient choice will be to plot the dimensionless ratio $k/\mu_2$.  The results then have the form
\eqn{kOscResult}{
{k_{\rm osc} \over \mu_2} = \pm \sqrt{ \left( \tilde{k}_{\rm osc} \over \mu_2 \right)^2} -{k_{\rm shift}\over \mu_2 } \,.
}
It is interesting to consider the expressions for $\tilde{k}_{\rm osc}^2/\mu_2^2$
as $\mu_R \to 0$:
\eqn{kOscLeft}{
\left(\tilde{k}_{\rm osc} \over \mu_2 \right)^2 = {q_2^2 - 4 m_1^2 \over 2} + (2 m_1^2 - 4 m_1 m_2 - q_2^2) \,\mu_R^2 + \sqrt{2} q_1 q_2\, \mu_R^3 + \ldots \,,
}
and as $\mu_R \to 1$,
\eqn{koscLeft}{
\left(\tilde{k}_{\rm osc} \over \mu_2 \right)^2 = {q_1^2 - 4 m_2^2 \over 4(1 - \mu_R)} + \left( {7 m_2^2 \over 2 }+ {q_1 q_2 \over \sqrt{2}} - 4 m_1 m_2 - {9 q_1^2 \over 8} \right)+ \ldots \,.
}
We see that to leading order, the existence of the oscillatory region as $\mu_R \to 0$ is determined by a competition between $q_2$ and $m_1$, while 
 the existence of the oscillatory region as $\mu_R \to 1$ is determined by a competition between $q_1$ and $m_2$.  In practice things are not quite so simple: we have $q_2^2 = 4 m_1^2$ for more than half our fermions, and $q_1^2 = 4 m_2^2$ holds for each and every one of them. Thus the existence of the oscillatory region is generally determined by the subleading terms, which we therefore include above.  The universal vanishing of $q_1^2 - 4 m_2^2$ also means that the expressions for $k_{\rm osc}$ are finite throughout the parameter space; this finiteness is another aspect that holds for our top-down fermions that would not be present in a generic bottom-up construction.

Fermi momenta are defined as values of $k \equiv k_F$ such that the source term in the near-boundary expansion \eno{BoundaryScalings} of the fermion vanishes:
\eqn{VanishingA}{
A(k_F) = 0 \,.
}
Fermi surfaces exist outside the oscillatory region.  Varying $\mu_R$  can cause a Fermi momentum to pass inside the oscillatory region, at which point the Fermi surface ceases to exist.

We will see that when a Fermi surface exists, the location of the Fermi momentum $k_F$ is often (but not always) right outside the oscillatory region.  Thus while we do not have an analytic formula for the value of $k_F$, we can gain some insight by using $k_{\rm osc}$ satisfying \eno{kOscResult} 
as a proxy for the Fermi surface location.  We will see there are several cases where $k_F \approx k_{\rm osc}$ holds quite well, but also a few where it does not.

Once a Fermi surface is located, the properties of nearby excitations may be studied.  Several of these depend solely on $\nu_{k_F}$ \cite{Faulkner:2009wj}. The retarded Green's function near the Fermi surface for small $\omega$  takes the form
\eqn{GreensFunction}{
G_R(k, \omega) \sim {h_1 \over k_\perp - {1 \over v_F} \omega - h_2 e^{i \gamma_{k_F}} (2 \omega)^{2 \nu_{k_F}}} \,,
}
with $h_1$, $h_2$ positive constants, $k_\perp \equiv k - k_F$ and the angle $\gamma_k$ defined by
\eqn{}{
\gamma_k \equiv \arg \left( \Gamma(-2 \nu_k) \left( e^{-2 \pi i \nu_k} - e^{-2 \pi  (q e)_{\rm eff}} \right) \right)\,.
}
For non-Fermi liquids, which have $\nu_{k_F} < 1/2$, we can ignore the Fermi velocity $v_F$ term as subleading.  The dispersion relation between the excitation energy $\omega_*$ and the momentum $k_\perp$ is then
\eqn{}{
\omega_* \sim (k_\perp)^z \,,
}
where the exponent is 
\eqn{}{
z \equiv {1 \over 2 \nu_{k_F}} \,.
}
Furthermore the residue $Z$ vanishes at the Fermi surface like
\eqn{}{
Z \sim (k_\perp)^{z-1} \,.
}
Finally, the ratio of the excitation width to its energy is given by
\eqn{Width}{
{\Gamma \over \omega_*} &= \tan \left( \gamma_{k_F} \over 2 \nu_{k_F} \right) \,, \quad \quad k_\perp>0 \,, \cr
&= \tan \left({ \gamma_{k_F} \over 2 \nu_{k_F} } - \pi z \right) \,, \quad \quad k_\perp<0 \,.
}
Different formulas hold for Fermi liquids, which have $\nu_{k_F} > 1/2$; as we shall see, all our Fermi surfaces are non-Fermi liquids, with one interesting special case approaching the marginal Fermi liquid at $\nu_{k_F} \to 1/2$.

\section{Fermi surfaces in 2+1-charge black holes}
\label{TwoPlusOneSec}

To obtain the locations $k_F$ of Fermi surfaces as a function of $\mu_R \equiv \mu_1/\mu_2$ for the various fermions, we numerically solve the decoupled second-order Dirac equation \eno{SecondOrderDirac} at $\omega =0$, beginning at the horizon where we impose the positive sign exponent in \eno{HorizonBCs} as a boundary condition, and searching for values of $k_F$ that cause the source term to vanish as \eno{VanishingA}.  We arbitrarily solve for the spinor component $\alpha = 1$; the other component $\alpha = 2$  has identical results with $k \to -k$.  We plot the values of $k_F/\mu_2$ vs.~$\mu_R$ for the Dirac equations with Fermi surfaces in the following figures, as well as $\nu_{k_F}$, $z$ and $\Gamma/\omega$ for each case.

A few general points before we consider each fermion in turn:

\begin{itemize}

\item{The fermions with asymptotic mass $m \to {3 \over 2L}$, which sit in the ${\bf 4}$ of $SO(6)$ and are dual to the operators Tr~$F_+ \lambda$, possess no Fermi surfaces.}

\begin{figure}
\begin{center}
\includegraphics[scale=0.35]{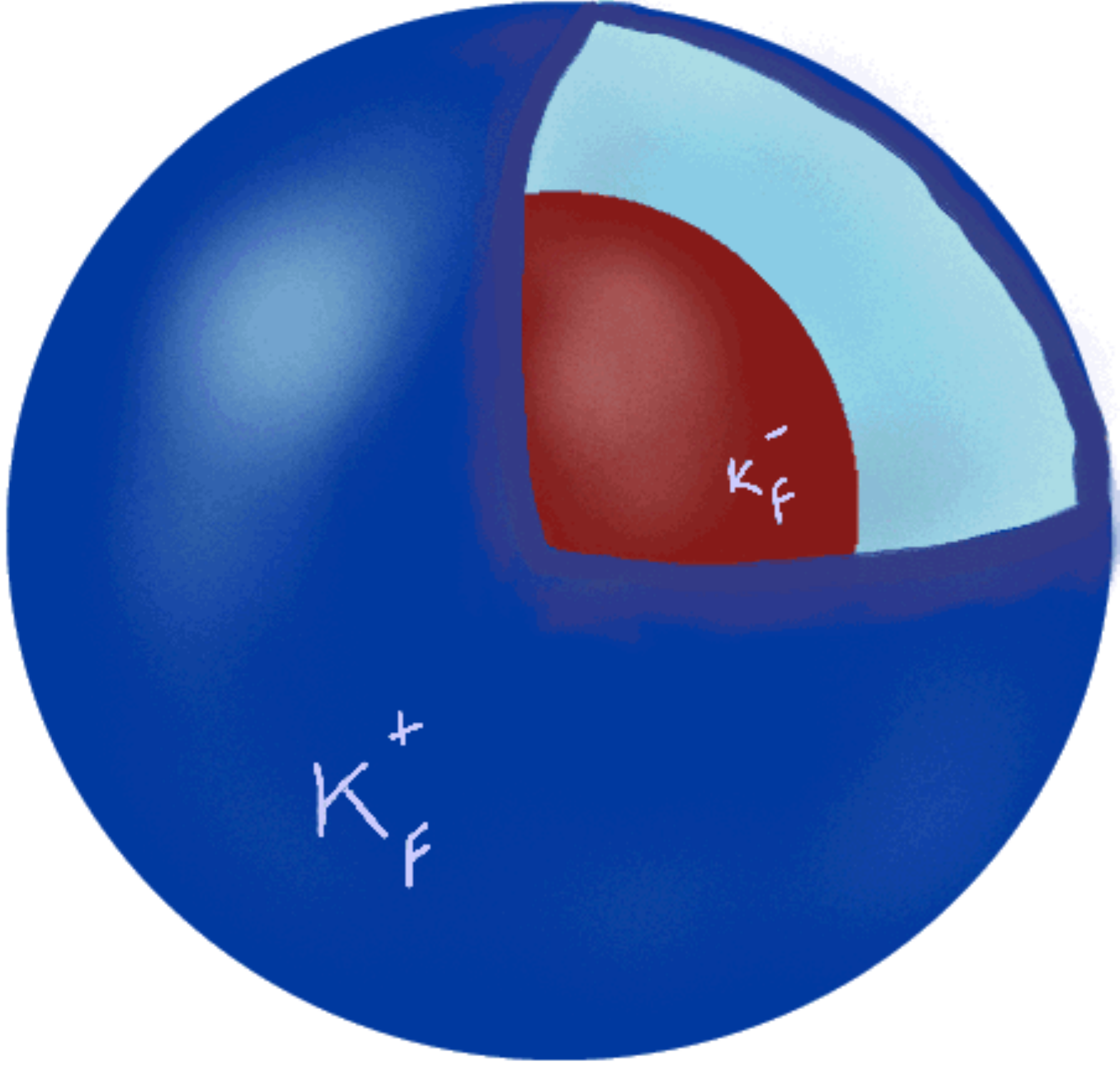} \quad \quad \quad \quad \quad \quad
\includegraphics[scale=0.3]{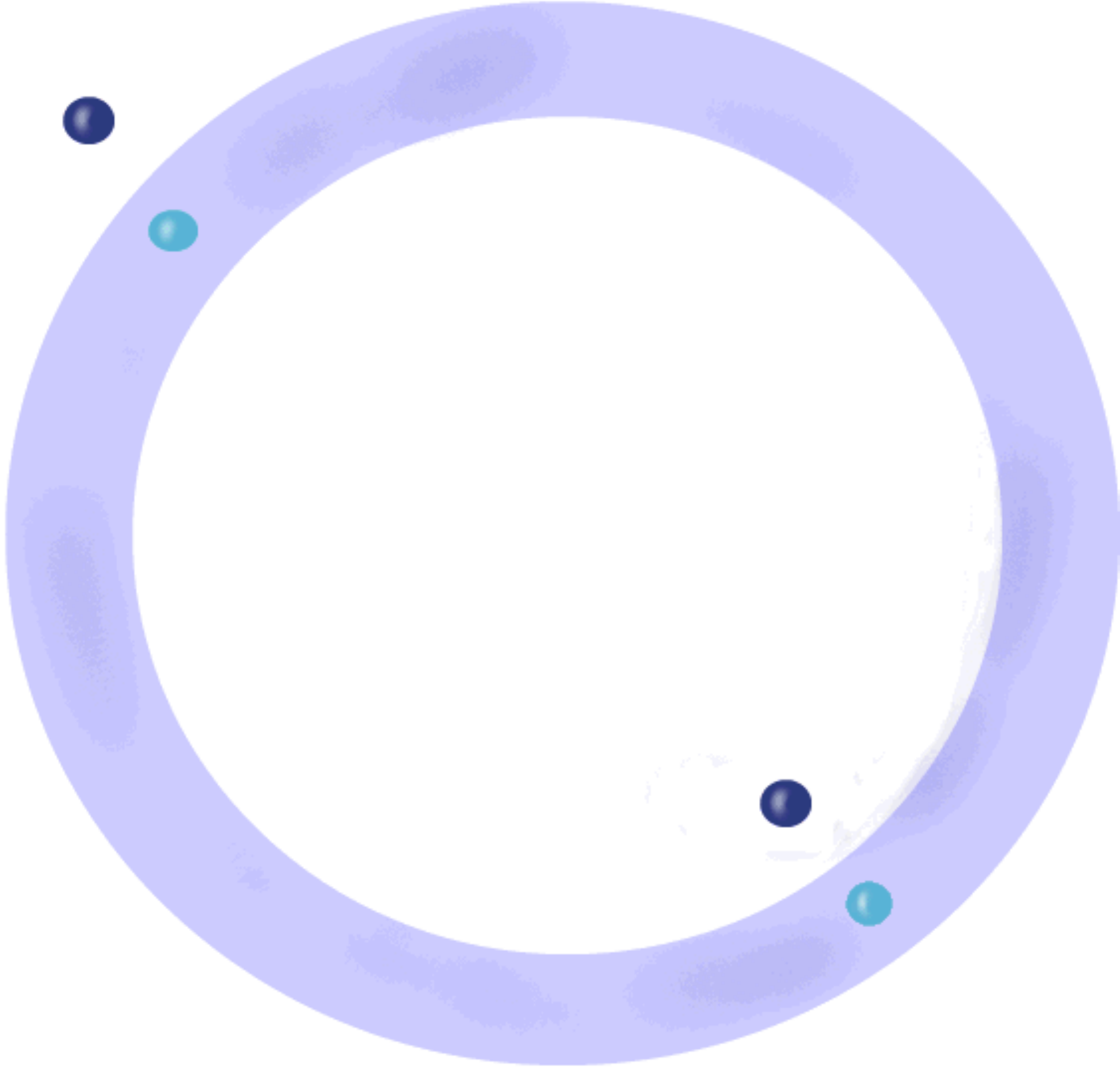}
\caption{Multiple fermi surface singularities correspond to nested spheres (left).  When the signs of $k_F$ are opposite and the excitation spectrum is the same at both surfaces, one interpretation is a thick shell (right).
\label{NestedSpheresFig}}
\end{center}
\end{figure}

\item{For the fermions with asymptotic mass $m \to {1 \over 2L}$,  sitting in the ${\bf 20}$ of $SO(6)$ and  dual to the operators Tr~$\lambda X$, the number of Fermi surfaces for each background can be classified by the total charge $q_3 \equiv q_1 + q_2$.  The two distinct equations with $q_3 = 5/2$ have two Fermi surfaces for at least part of the range of $\mu_R$; the three distinct equations with $q_3 = 3/2$ have one Fermi surface for at least part of the range of $\mu_R$; and those with $q_3 = 1/2$ have none.  This is in harmony with a general expectation that fermions with greater charge are more likely to form Fermi surfaces.
 }

\item{The Fermi surfaces have $0 < \nu_{k_F} < 1/2$, indicating they are all non-Fermi liquids.}

\item{The Fermi surfaces for the most part exist close to the boundary of the oscillatory region; in each plot we include the values of $k_{\rm osc}$ and shade the oscillatory region.  As $\mu_R$ is varied, $k_F$ may pass below $k_{\rm osc}$, at which point the Fermi surface ceases to exist.  As the Fermi surface approaches the oscillatory region we have $\nu_{k_F} \to 0$, implying the dispersion relation exponent $z \to \infty$ as the system moves further into the deeply non-Fermi regime.}

\item{At the 3-charge black hole point, denoted as 3QBH, the two Dirac equations $q_3 = 5/2$ coincide, as do the three equations with $q_3 = 3/2$.  Thus the values of $k_F$ and $k_{\rm osc}$ must match for fermions of the same type, as indeed they do. The $q_3 = 5/2$ modes at the 3-charge black hole point are the fluctuations studied in \cite{DeWolfe:2011aa}, and we match the results found there; different radial coordinates and conventions for the sign of the gauge fields mean that $k_{\rm here}= -\sqrt{3} k_{\rm there}$.}

\item{When the spectrum of a fermion with charges $(q_i, p_i)$ contains a Fermi surface singularity at $k_F$, the conjugate (antiparticle) equation with charges $(-q_i, -p_i)$ will have a Fermi surface singularity at $-k_F$ (see equation~\eno{SignFlips}); allowing $k$ to point in the $y$ and $z$ directions will complete these antipodal points into a full three-dimensional Fermi sphere.  For the examples where multiple values of $k_F$ are observed, this corresponds to nested Fermi spheres (see figure~\ref{NestedSpheresFig}).
}

\item{As noted in equation~\eno{Width}, one can examine excitations around Fermi surfaces with $k_\perp \equiv k- k_F >0$, corresponding to particles, and $k_\perp < 0$, corresponding to holes; such excitations should be on the physical sheet of the complex $\omega$-plane \cite{Faulkner:2009wj}.  We found that excitations with $k_\perp < 0$ exist on the physical sheet for small $\omega$, but those with $k_\perp>0$ do not, a manifestation of particle/hole asymmetry.  In the following we plot $\Gamma/\omega$ for $k_\perp < 0$.  The role of particles and holes is exchanged for the conjugate equation with charges $(-q_i, -p_i)$.}

\end{itemize}

We turn now to a discussion of the results for each of the five distinct Dirac equations for which we find Fermi surface singularities.

\bigskip\bigskip
\noindent
{\bf Case 1: $\chi^{({1 \over 2} , {3 \over 2}, {1 \over 2})}$ and $\chi^{({1 \over 2} , {1 \over 2}, {3 \over 2})}$, dual to Tr $\lambda_1 Z_2$ and Tr $\lambda_1 Z_3$}

\begin{figure}
\begin{center}
\includegraphics[scale=1.0]{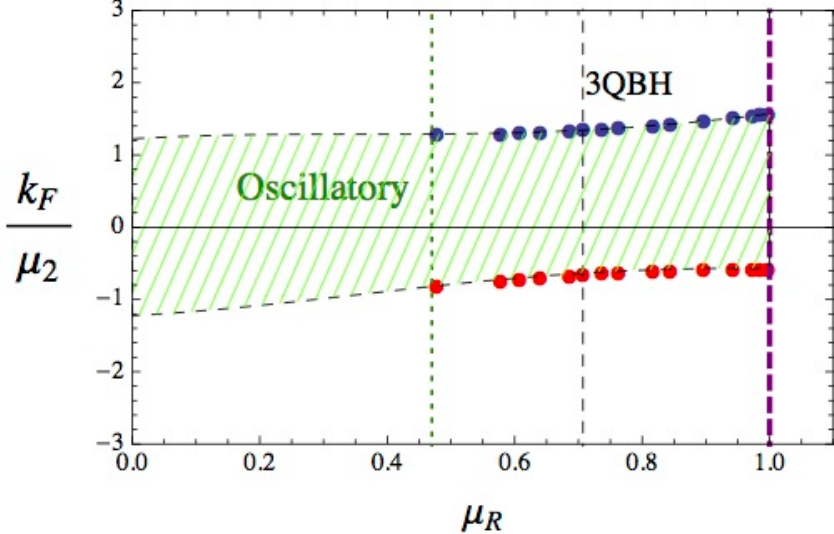}
\caption{The values of $k_F/\mu_2$ for case~1,
Tr $\lambda_1 Z_2$ and Tr $\lambda_1 Z_3$.
\label{FSPlot131}}
\end{center}
\end{figure}

\begin{figure}
\begin{center}
\includegraphics[scale=0.8]{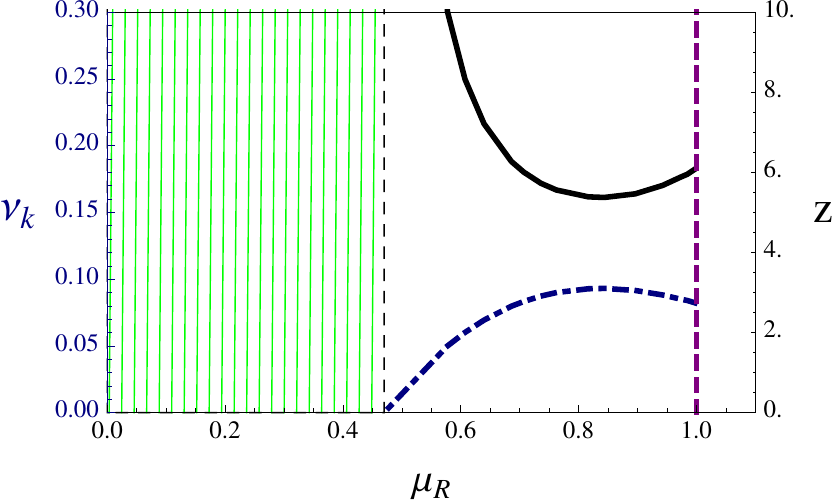} \quad \quad
\includegraphics[scale=0.7]{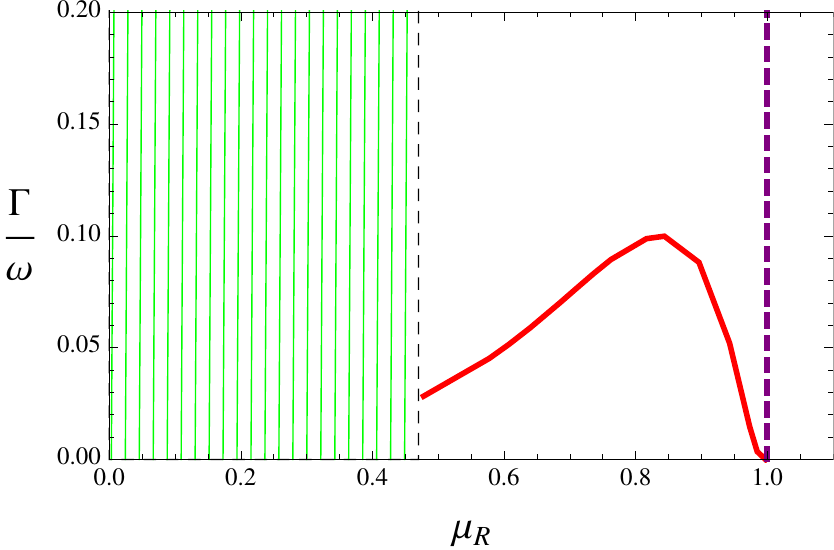}
\caption{The values of $\nu_{k_F}$, $z$ and $\Gamma/\omega$ for case~1,
Tr $\lambda_1 Z_2$ and Tr $\lambda_1 Z_3$.
\label{ParamsPlot131}}
\end{center}
\end{figure}

\smallskip
\noindent
These modes have $(q_1, q_2) = ({1 \over 2}, 2)$, $(p_1, p_2) = ({1 \over 4}, 0)$ and $(m_1, m_2) = ({1 \over 2}, -{1 \over 4})$; $k_F$ is given in figure~\ref{FSPlot131} and $\nu_{k_F}$, $z$ and $\Gamma/\omega$ in figure~\ref{ParamsPlot131}.

The oscillatory region has the shape of a wavy band, as $\tilde{k}_{\rm osc}/\mu_2$ approaches a constant value in either limit, while $k_{\rm shift}$ approaches zero at $\mu_R \to 0$ but a finite value at $\mu_R \to 1$.  This fermion is one of two with the largest total charge $q_5 = 5/2$, and has two Fermi surfaces for larger values of $\mu_R$; these track the oscillatory region boundary very closely.  At the 3-charge point these fermions reduce to the results of \cite{DeWolfe:2011aa}. The Fermi surfaces disappear into the oscillatory region at around $\mu_R \approx 0.47$.

The values of $\nu_k$ are small throughout the range, resulting in a scaling exponent $z$ for the excitations that never gets smaller than $z \gtrsim 5$, and which (as it must) grows without bound as the Fermi momenta approach the oscillatory region.  The ratio $\Gamma/\omega$ of excitation width also stays small, being bounded above by $\Gamma/\omega \approx 1/10$ and going to zero as $\mu_R \to 1$; thus in this limit the would-be quasiparticle excitations become more and more stable.

In the full three-dimensional $k$-space, these two Fermi surface singularities are completed into nested spheres.  The excitation spectrum near both surfaces is the same; since the two $k_F$ solutions are of opposite sign, this implies that the excitations inside the outer sphere match those outside the inner sphere.  An interpretation of the two Fermi spheres is then that there is a ``thick shell" of occupied states in between them, with unoccupied states both on the outside and the inside of the pair (see Figure~\ref{NestedSpheresFig}).  Put another way, there is a Fermi surface of antiparticles inside the surface of particles, canceling out in the overlap.  It is also possible that the Fermi surfaces are associated to distinct  would-be quasiparticles, but they must have opposite sign.

\bigskip
\noindent
{\bf Case 2: $\chi^{({3 \over 2} , {1 \over 2}, {1 \over 2})}$, dual to Tr $\lambda_1 Z_1$}

\begin{figure}
\begin{center}
\includegraphics[scale=1.0]{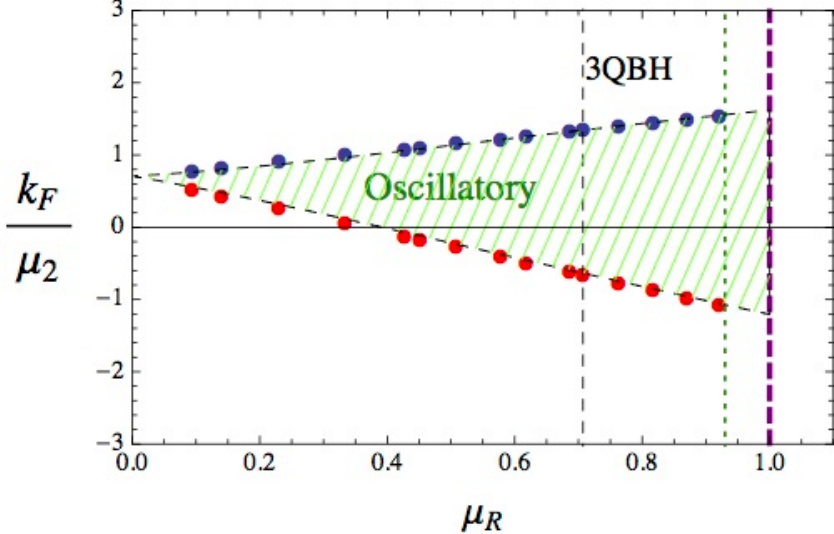}
\caption{The values of $k_F/\mu_2$ for case~2,
Tr $\lambda_1 Z_1$.
\label{FSPlot311}}
\end{center}
\end{figure}

\begin{figure}
\begin{center}
\includegraphics[scale=0.8]{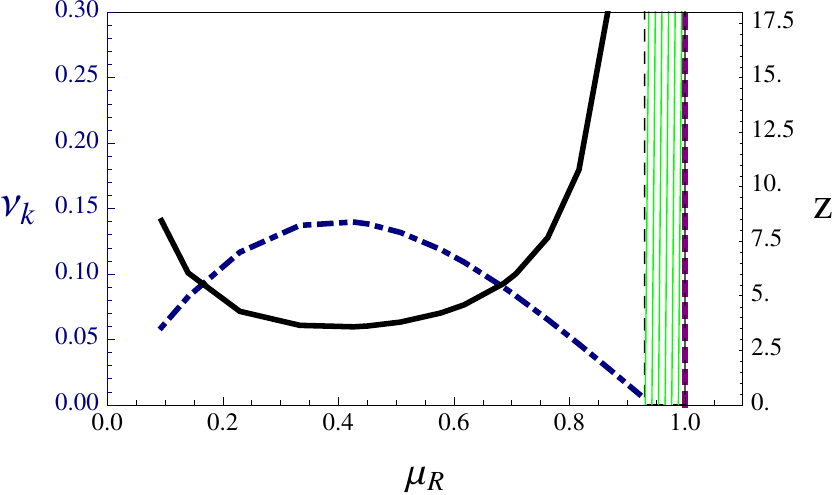} \quad \quad
\includegraphics[scale=0.7]{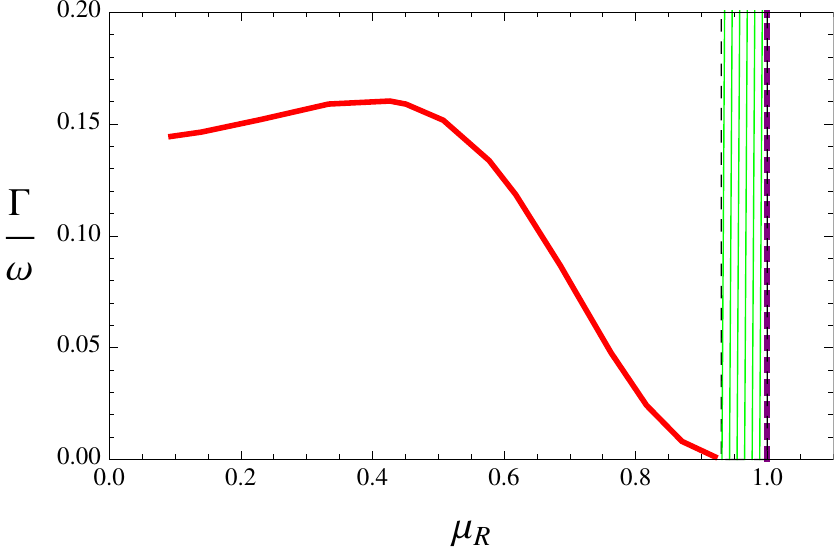}
\caption{The values of $\nu_{k_F}$, $z$ and $\Gamma/\omega$ for case~2,
Tr $\lambda_1 Z_1$.
\label{ParamsPlot311}}
\end{center}
\end{figure}

\smallskip
\noindent
This mode has $(q_1, q_2) = ({3 \over 2}, 1)$, $(p_1, p_2) = (-{1 \over 4}, {1 \over 2})$ and $(m_1, m_2) = (-{1 \over 2}, {3 \over 4})$; $k_F$ is given in figure~\ref{FSPlot311} and $\nu_{k_F}$, $z$ and $\Gamma/\omega$ in figure~\ref{ParamsPlot311}.

As in  case 1, both $\tilde{k}_{\rm osc}$ and $k_{\rm shift}$ approach finite values at $\mu_R \to 1$.  At $\mu_R \to 0$, however, the vanishing of $q_2^2 - 4 m_1^2$ means that $\tilde{k}_{\rm osc}$ goes to zero, while $k_{\rm shift}$ reaches a finite value. Thus both branches approach $k_{\rm osc}/\mu_2 \to 1/\sqrt{2}$; the separation goes as ${\cal O}(\mu_R^2)$.

This is the other fermion with largest total charge $q_3 = 5/2$, and it also has two Fermi surfaces for much of the range of $\mu_R$, again tracking the oscillatory region very closely.  Unlike case~1, here the Fermi surfaces exist as far as can be determined for arbitrarily small $\mu_R$, while they disappear into the oscillatory region on the right side of the plot, around $\mu_R \approx 0.93$.
At the 3-charge point we coincide with case~1 and the results of \cite{DeWolfe:2011aa}. 
The values of $\nu_k$ are again small, though the maximum is slightly greater than the previous case, and the exponent reaching a minimum around $z \gtrsim 3.5$, with  $\Gamma/\omega \lesssim 1/15$.

Interestingly, in this instance one of the Fermi momenta passes through zero, while the other does not.  To the right of the zero, this case resembles case 1, where the two values of $k_F$ have opposite sign and one interpretation is an occupied shell in momentum space.  To the left of the zero, however, the two Fermi singularities have the same sign of $k_F$, implying the same excitations outside each Fermi surface; the nested spheres now have the same sign excitations.

\bigskip
\noindent
{\bf Case 3: $\bar\chi^{(-{1 \over 2} , {3 \over 2}, {1 \over 2})}$ and $\bar\chi^{(-{1 \over 2} , {1 \over 2}, {3 \over 2})}$, dual to Tr $\overline\lambda_2 Z_2$ and Tr $\overline\lambda_2 Z_3$}

\begin{figure}
\begin{center}
\includegraphics[scale=1.0]{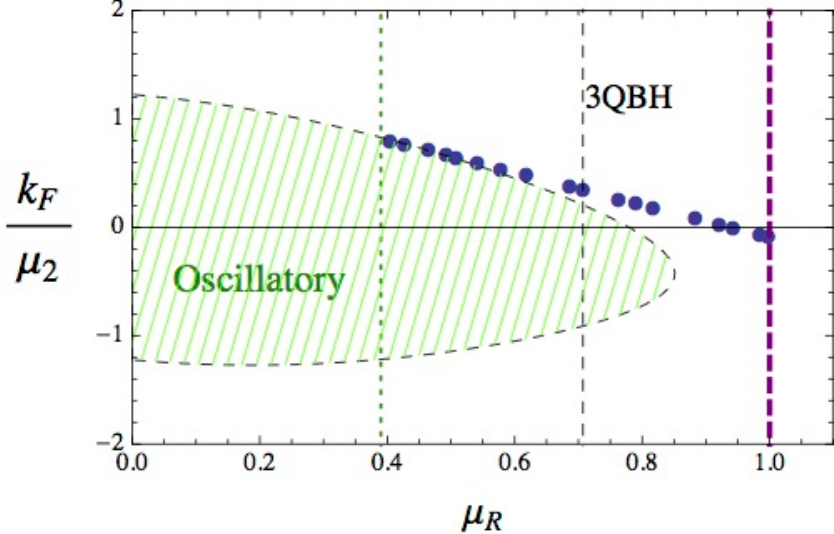}
\caption{The values of $k_F/\mu_2$ for case~3,
Tr $\overline\lambda_2 Z_2$ and Tr $\overline\lambda_2 Z_3$.
\label{FSPlotm131}}
\end{center}
\end{figure}

\begin{figure}
\begin{center}
\includegraphics[scale=0.8]{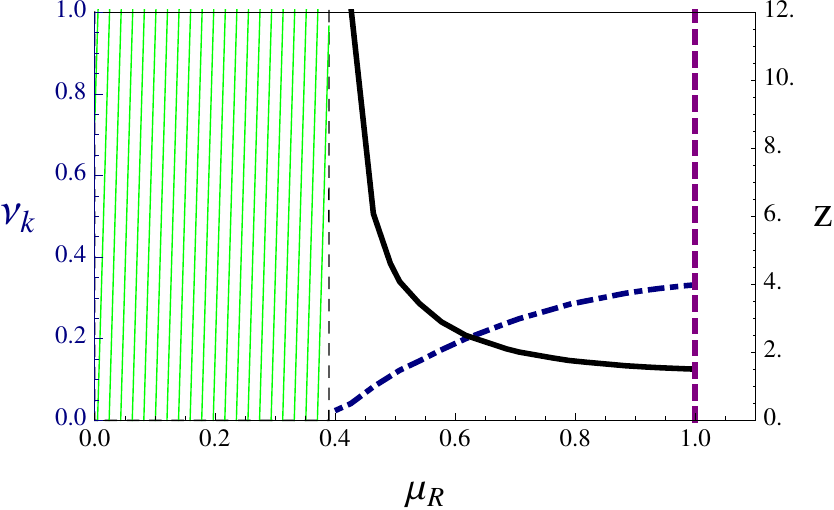} \quad \quad
\includegraphics[scale=0.7]{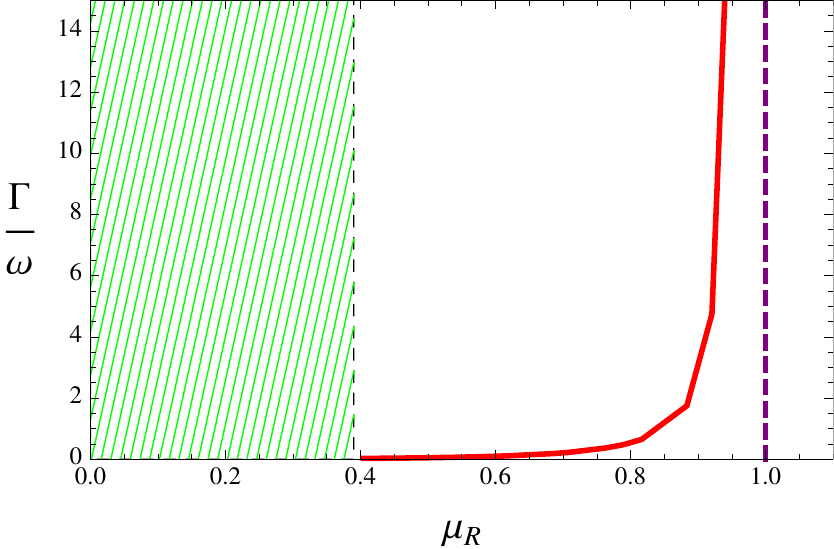}
\caption{The values of $\nu_{k_F}$, $z$ and $\Gamma/\omega$ for case~3,
Tr $\overline\lambda_2 Z_2$ and Tr $\overline\lambda_2 Z_3$.
\label{ParamsPlotm131}}
\end{center}
\end{figure}

\smallskip
\noindent
These modes have $(q_1, q_2) = (-{1 \over 2}, 2)$, $(p_1, p_2) = (-{1 \over 4}, 0)$ and $(m_1, m_2) = ({1 \over 2}, -{1 \over 4})$; $k_F$ is given in figure~\ref{FSPlotm131} and $\nu_{k_F}$, $z$ and $\Gamma/\omega$ in figure~\ref{ParamsPlotm131}.

Here the oscillatory region is a band as $\mu_R \to 0$, with vanishing $k_{\rm shift}$ there due to $p_2 = 0$, similar to case 1.  Before $\mu_R$ can reach 1, however, the effective mass term dominates over the effective coupling and $\tilde{k}_{\rm osc}$ ceases to exist.  Hence for black holes with $0.85 \lesssim \mu_R < 1$, no oscillatory region exists for any $k$.

This fermion has total charge $q_3 = 3/2$, and there is one Fermi surface for most values of $\mu_R$.  The Fermi surface emerges from the oscillatory region close to $\mu_R \approx 0.4$, and above this $k_F$ tracks the value of $k_{\rm osc}$ for a while; however they soon separate and the fermi surface continues to exist even as $\mu_R \to 1$, once the oscillatory region has disappeared.  Notice that $k_F$ goes through zero: this indicates that the Fermi surface shrinks to zero size and then re-expands with antiparticle excitations.
The exponent $z$ as usual diverges as the Fermi surface enters the oscillatory region; it finds its minimum at $\mu_R \to 1$, where it passes below $z = 2$.  The excitation widths over energies also diverge before the maximum $\mu_R$ value is reached, and the corresponding pole moves off the physical sheet on the complex $\omega$ plane.

\bigskip
\noindent
{\bf Case 4: $\bar\chi^{({1 \over 2} , -{1 \over 2}, {3 \over 2})}$ and $\bar\chi^{({1 \over 2} , {3 \over 2}, -{1 \over 2})}$, dual to Tr $\overline\lambda_3 Z_3$ and Tr $\overline\lambda_4 Z_2$.}

\begin{figure}
\begin{center}
\includegraphics[scale=1.0]{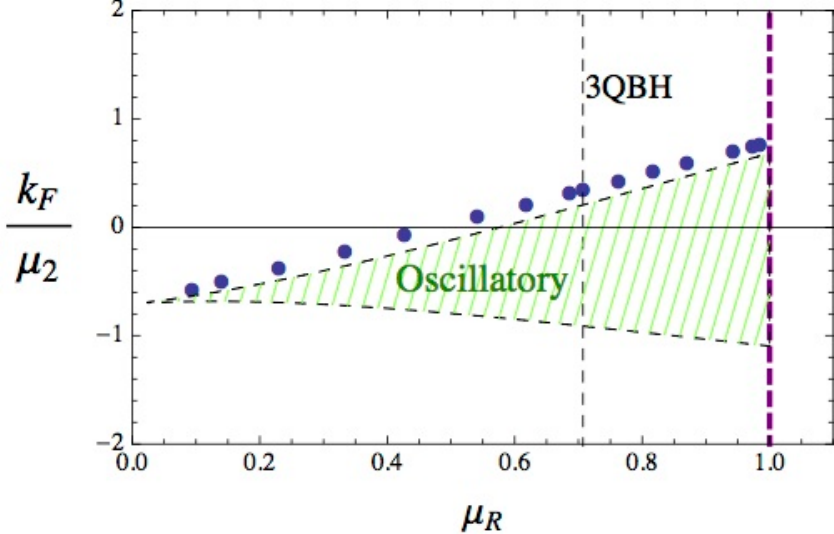}
\caption{The values of $k_F/\mu_2$ for case~4,
Tr $\overline\lambda_3 Z_3$ and Tr $\overline\lambda_4 Z_2$.
\label{FSPlot1m13}}
\end{center}
\end{figure}

\begin{figure}
\begin{center}
\includegraphics[scale=0.8]{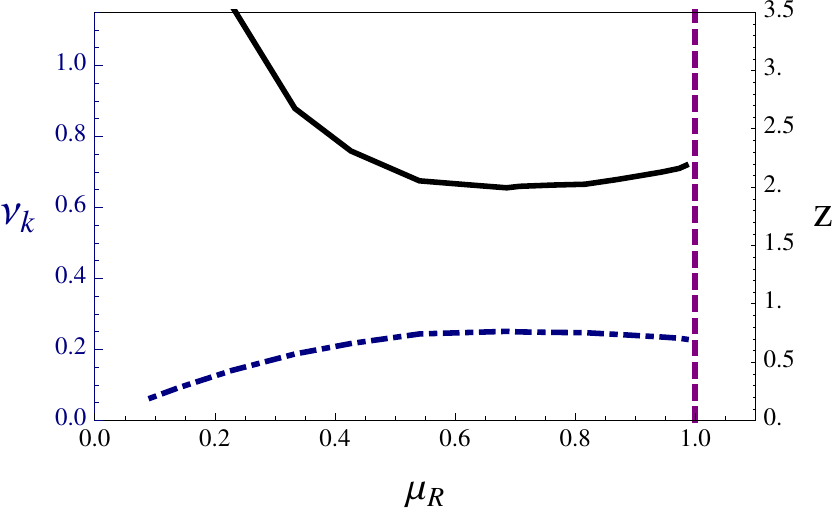} \quad \quad
\includegraphics[scale=0.7]{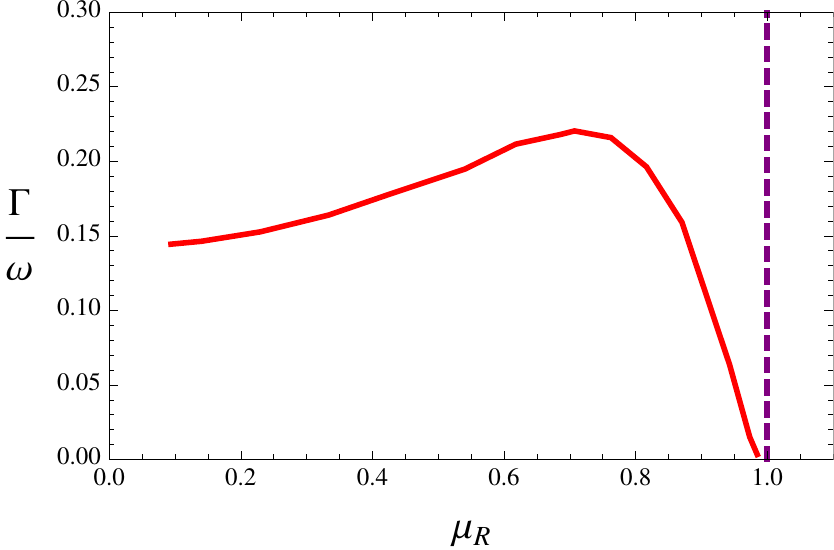}
\caption{The values of $\nu_{k_F}$, $z$ and $\Gamma/\omega$ for case~4,
Tr $\overline\lambda_3 Z_3$ and Tr $\overline\lambda_4 Z_2$.
\label{ParamsPlot1m13}}
\end{center}
\end{figure}

\smallskip
\noindent
These modes have $(q_1, q_2) = ({1 \over 2}, 1)$, $(p_1, p_2) = ({1 \over 4}, -{1 \over 2})$ and $(m_1, m_2) = ({1 \over 2}, -{1 \over 4})$; $k_F$ is given in figure~\ref{FSPlot1m13} and $\nu_{k_F}$, $z$ and $\Gamma/\omega$ in figure~\ref{ParamsPlot1m13}.

The behavior in this fermion is similar to case~2, with $\tilde{k}_{\rm osc}$ vanishing at $\mu_R \to 0$; in fact the charges and mass values conspire so that the first two terms in the $\mu_R \to 0$ expansion vanish, leaving a cuspy ${\cal O}(\mu_R^3)$ deviation away from the limiting value $k_{\rm osc}/\mu_2 \to -1/\sqrt{2}$.

The total charge is $q_3 = 3/2$, and again there is a single Fermi surface.  For this example the Fermi surface never enters into the oscillatory region that we can discern, although it tracks it rather closely.  The Fermi momentum again goes through zero, indicating a Fermi surface of particles transitioning to one of antiparticles.
The exponent $z$ again reaches a minimum around $z \approx 2$, and the ratio $\Gamma/\omega$ has a maximum around $\Gamma/\omega \lesssim 0.23$, going to zero as $\mu_R \to 1$.

\bigskip
\noindent
{\bf  Case 5: $\bar\chi^{({3 \over 2} , -{1 \over 2}, {1 \over 2})}$ and $\bar\chi^{({3 \over 2} , {1 \over 2}, -{1 \over 2})}$, dual to Tr $\overline\lambda_3 Z_1$ and Tr $\overline\lambda_4 Z_1$.}

\begin{figure}
\begin{center}
\includegraphics[scale=1.0]{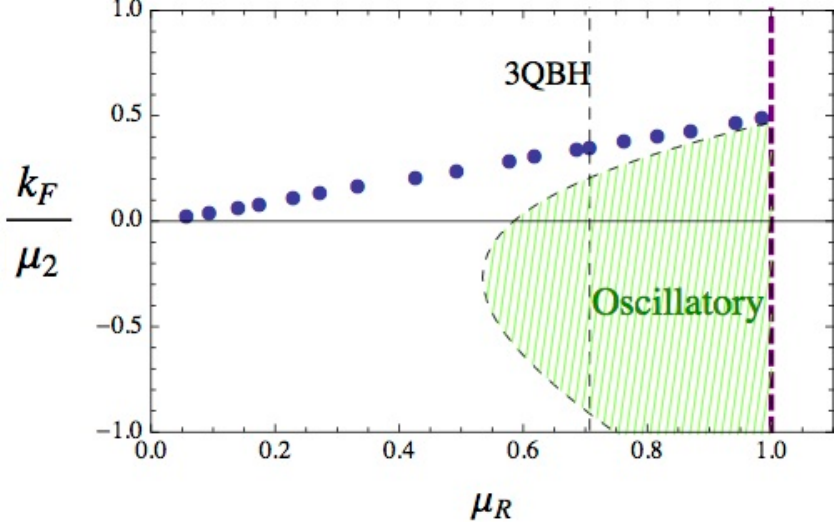}
\caption{The values of $k_F/\mu_2$ for case~5,
Tr $\overline\lambda_3 Z_1$ and Tr $\overline\lambda_4 Z_1$.
\label{FSPlot3m11}}
\end{center}
\end{figure}

\begin{figure}
\begin{center}
\includegraphics[scale=0.8]{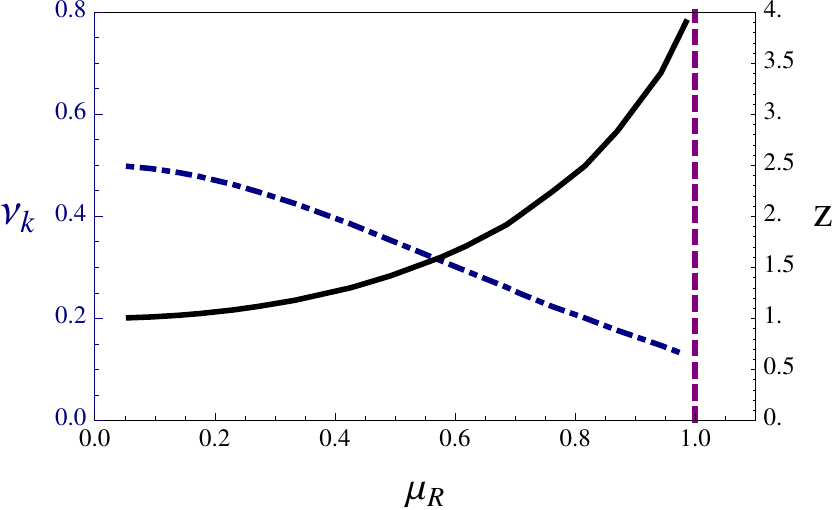} \quad \quad
\includegraphics[scale=0.7]{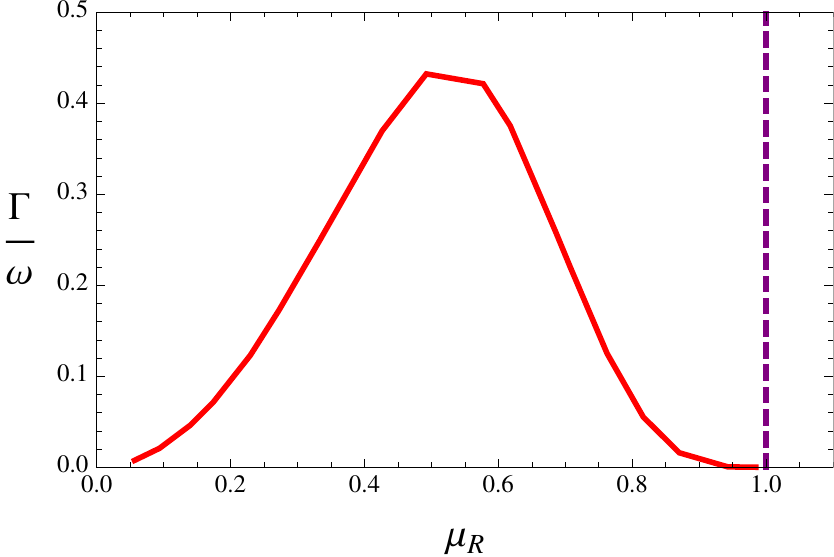}
\caption{The values of $\nu_{k_F}$, $z$ and $\Gamma/\omega$ for case~5,
Tr $\overline\lambda_3 Z_1$ and Tr $\overline\lambda_4 Z_1$.
\label{ParamsPlot3m11}}
\end{center}
\end{figure}

\smallskip
\noindent
These modes have $(q_1, q_2) = ({3 \over 2}, 0)$, $(p_1, p_2) = (-{1 \over 4}, 0)$ and $(m_1, m_2) = (-{1 \over 2}, {3 \over 4})$; $k_F$ is given in figure~\ref{FSPlot3m11} and $\nu_{k_F}$, $z$ and $\Gamma/\omega$ in figure~\ref{ParamsPlot3m11}.

In this case the oscillatory region vanishes for $\mu_R \lesssim 0.53$ as the effective mass overwhelms the effective electric coupling.
The effective charge is $q_3 = 3/2$, and there is again one Fermi surface; as with the other two $q_3 = 3/2$ cases, the Fermi surface is above the oscillatory region.  As $\mu_R \to 1$, the Fermi momentum appears to meet the oscillatory region.  It departs from the oscillatory region before $k_{\rm osc}$ ceases to exist, as in case~3, and interestingly, appears to head towards $k_F /\mu_2\to 0$.

This fermion is remarkable because it has $q_2 =0$ and $p_2= 0$; it is unaware of the $A_\mu$ gauge field.
The effective electric coupling $(qe)_{\rm eff}^2$ given in \eno{GaugeAndMass} in general has the leading behavior as $\mu_R \to 0$ of $(qe)_{\rm eff}^2 \to q_2^2/4$; with $q_2 =0$, the subleading behavior does not kick in until $(qe)_{\rm eff}^2 \to 9 \mu_R^6/8$.  The effective electric coupling is thus very close to zero for much of the range of $\mu_R$.

The width ratio $\Gamma/\omega$ reaches a maximum of $\Gamma/\omega \approx 0.45$ in the middle of the range of $\mu_R$, dropping off to zero on either side.  Perhaps most interestingly, $\nu_{k_F}$ steadily climbs as $\mu_R \to 0$, approaching $\nu_{k_F} \to 1/2$; this is the limit of the marginal Fermi liquid, where the non-Fermi liquid crosses over to an ordinary Fermi liquid.  Correspondingly, the exponent $z$ approaches $z \to 1$.

It is tempting to look for an explanation for such limiting behavior.  As previously noted, the limit $\mu_R \to 0$ is related to the BPS Coulomb branch solution of the 1-charge black hole, although the order of limits is subtle as $\Phi_2$ remains nonzero.  However, this fermion --- which we call the ``marginal Fermi liquid" fermion --- has $q_2 = p_2 =0$, and hence is unaware of $\Phi_2$.  Thus, as $\mu_R \to 0$, its Dirac equation becomes a Dirac equation in the Coulomb branch background.  In the next section, we derive these results analytically by matching onto the Coulomb branch solution.

\section{The 1-charge black hole and the marginal Fermi liquid fermion}
\label{OneChargeSec}

In this section we explore the BPS Coulomb branch solution that is the $r_H \to 0$ limit of the 1-charge black hole.  We first describe how the $\mu_R \to 0$ limit of the``marginal Fermi liquid" fermion is controlled by the infinitesimal-$k$ limit of the Coulomb branch solution.  We then describe the solutions for all our fermionic modes at finite $k$, which like other related modes explored in the literature manifest a continuous distribution over a gap. We then describe conductivities in this background, and argue that the $a_\mu$ gauge field is superconducting while $A_\mu$ is insulating, as well as considering the extension of the conductivities into the 2+1-charge backgrounds, where the superconducting delta function in the conductivity broadens to a Drude peak.

\subsection{Derivation of the marginal Fermi liquid limit}

The fermions $\bar\chi^{({3 \over 2}, -{1 \over 2}, {1 \over 2})}$ and $\bar\chi^{({3 \over 2}, {1 \over 2}, -{1 \over 2})}$, referred to as case 5 in the previous section, possess 
the distinction of having $q_2 = p_2 = 0$.  As a result, these modes are completely insensitive to the $\Phi_2$ gauge field, and as described before there is a limit where the Dirac equation in the 2+1-charge extremal background approaches that of the BPS Coulomb branch background, where things simplify.  As we shall see, the existence of a Fermi surface approaching the marginal Fermi liquid value of $\nu_{k_F} = 1/2$ can be determined analytically there.

\begin{figure}
\begin{center}
\includegraphics[scale=1.0]{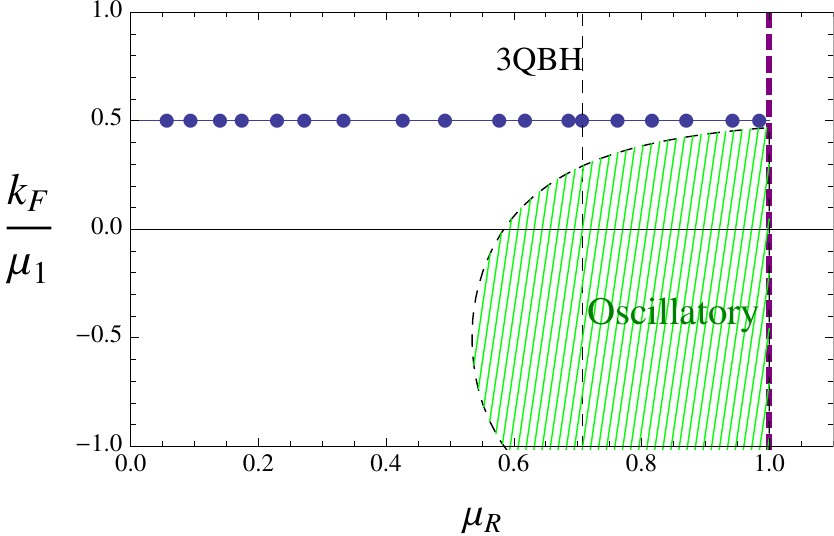}
\caption{The values of $k_F/\mu_1$ for case~5, the MFL fermions
Tr $\overline\lambda_3 Z_1$ and Tr $\overline\lambda_4 Z_1$.
\label{FSPlot3m11mu1}}
\end{center}
\end{figure}

It is interesting to consider the behavior of $k_F/\mu_1$, instead of $k_F/\mu_2$, for solutions of this Dirac equation.  For the other cases, $\mu_1 \to 0$ with $\mu_2$ fixed is the straightforward realization of $\mu_R \to 0$, and $k_F/\mu_2$ and $k_{\rm osc}/\mu_2$ remain finite in that limit. For the MFL fermion, however, it does not feel $\mu_2$ and so allowing $\mu_2 \to \infty$ with $\mu_1$ fixed is a reasonable limit.  As can be seen in figure~\ref{FSPlot3m11mu1}, in this normalization we have the result
\eqn{}{
k_F \approx {1 \over 2} \mu_1 \,,
}
over the entire range of $\mu_R$; the line $k_F/\mu_1 = 1/2$ is shown explicitly.  As we shall see, in the $\mu_R \to 0$ limit this relationship can be shown to be exact.  It is possible it is exact for the entire range of $\mu_R$, but we will not show this.

The 2+1-charge extremal backgrounds, having eliminated $Q_2$ to impose extremality, depend on the length scales $r_H$ and $Q_1$.  The limit approaching the Coulomb branch solution is $r_H \to 0$ with $Q_1$ finite; thus there is a separation of scales, $r_H \ll Q_1$.  Studying Fermi surfaces requires setting the right boundary conditions at the horizon.  Consequently, we will first study the region ``zoomed in" around $r_H$; keeping $r_H$ finite means $Q_1 \to \infty$.  We keep $k$ finite in this limit as well, but it drops out of the equations.  The Dirac equation for all four modes reduces to
\eqn{}{
\Psi''(r) + {r^2 + r_H^2 \over r (r^2 - r_H^2)} \Psi'(r) - {r^2 \over (r^2 - r_H^2)^2} \Psi(r) \,,
}
which has the solutions
\eqn{}{
\Psi_{\pm 1/2} (r) = (r^2 - r_H^2)^{\pm {1\over 2}} \,.
}
Near the horizon, these approach
\eqn{}{
\Psi_{\pm1/2}(r \to r_H) \to \left[ 2 r_H (r - r_H) \right]^{\pm {1\over 2}} \,,
}
corresponding to a value of $\nu_k = 1/2$.  Indeed we may verify that the formula \eno{nuk} for $\nu_k$ evaluated for this fermion in the large $Q_1$ limit gives $\nu_k \to 1/2$.

Thus to pick the ingoing/regular boundary condition, we pick the positive exponent $\nu_k = + 1/2$.  Meanwhile, at large $r$ these solutions become
\eqn{}{
\Psi_{+1/2}(r \to \infty) \to r + {\cal O}(r_H^2/ r) \,, \quad \quad
\Psi_{-1/2} (r \to \infty) \to {1 \over r} \,.  
}
Thus the regular solution at the horizon goes like $r$ far away from the horizon.

At values of $r$ much greater than $r_H$ one finds the scale $Q_1$, which we scaled away in the previous limit.  Let us now ``zoom in" on this region, taking $Q_1$ finite and $r_H \to 0$.
Here we approach the Coulomb branch solution (modulo the behavior of $\Phi_2$, which this fermion is insensitive to) where the Dirac equation simplifies.  Since we had $k$ of order $r_H$ in the inner region, we will do the same in this outer region, setting
\eqn{}{
k \equiv k_p r_H \,,
}
and holding $k_p$ finite in the limit.

To find a Fermi surface we study the $\Psi_{1 +}$ mode (as usual $\Psi_{2 +}$ has the same solutions with $k \to -k$), and the solutions are
\eqn{OuterRegionSolns}{
\Psi_{1 +} = A \, {(r^2 + Q_1^2)^{3/4} \over r}  + B \, {Q_1^2 (1 - k_p) + r^2 (3 - k_p) \over 2 r (r^2 + Q_1^2)^{5/4}} \,,
}
where $A$ and $B$ are constants.  We have chosen these names for the constants for good reason, for in the large-$r$ limit we find
\eqn{}{
\Psi_{1 +}(r \to \infty) \to A \, r^{1/2} + B \, {3 - k_p \over 2} r^{-3/2} \,,
}
which are the proper boundary scalings for $mL = 1/2$ according to \eno{BoundaryScalings}; $A$ is the source term.  For a Fermi surface, $A$ will vanish when infalling boundary conditions are imposed.

To determine what is implied by infalling boundary conditions, we match the small-$r$ limit of this outer region to the large-$r$ limit of the inner region just discussed, patching the solution together between the scales $r_H$ and $Q_1$.  The small-$r$ limit of the outer region solutions \eno{OuterRegionSolns} is
\eqn{}{
\Psi_{1 +}(r \to 0) \to A \left( {Q_1^{3/2} \over r} + {3 r \over 4 Q_1^{1/2}}  \right)+ B \left( {1 - k_p \over 2 Q_1^{1/2} r} + {(7 + k_p) r \over 8 Q_1^{5/2}} \right) \,.
}
To match to the $\nu_k = +1/2$ inner region solution, this must go like $r$; thus infalling boundary conditions requires
\eqn{}{
A = {(k_p - 1) B \over 2 Q_1^2} \,.
}
Thus for generic $k_p$, there is a nonzero source term.  However, for the special value
\eqn{}{
k_p = 1 \quad \to \quad k =  r_H \,,
}
we find $A =0$ when we impose infalling boundary conditions, and thus there is a Fermi surface.  Moreover at $\mu_R \to 0$, we have
\eqn{}{
{r_H \over L^2} (\mu_R \to 0) \to {\mu_1 \over 2} \,.
}
Thus we expect that as $Q_1/r_H \to \infty$, which corresponds to $\mu_1/\mu_2 \to 0$, there will be a Fermi surface in the family of extremal 2+1-charge black holes with parameters approaching
\eqn{}{
k_F \to r_H = {\mu_1 \over 2} \,, \quad \quad \nu_{k_F} \to {1 \over 2} \,.
}
This exactly matches the results obtained previously.  Figure~\ref{FSPlot3m11mu1} suggests that the relation $k_F \approx \mu_1/2$ continues to hold away from $\mu_R = 0$, although $k_F = r_H$ no longer will.

\subsection{Finite momentum fermion solutions for the Coulomb branch solution}
\label{RGFlowSec}

In the above analysis, the infinitesimal-$k$ limit of fermion fluctuations in the Coulomb branch background was matched on to the limit of the 2+1-charge extremal black holes for the marginal fermi liquid fermion.  One may also consider the finite-$k$ fermion fluctuations, and in the true Coulomb branch background rather than the 2+1-charge extremal limit, which we discuss here.

As discussed, this  background is a zero-temperature, zero-chemical potential renormalization group flow geometry, referred to as the ``$n=2$" Coulomb branch flow in \cite{Freedman:1999gk}, where the fluctuation equation for the transverse traceless graviton modes (which appear as a free scalar) were solved.  The coupled system of the scalar and the trace of the graviton was considered in \cite{DeWolfe:2000xi}, \cite{Arutyunov:2000rq}, and then several other modes, including the spin-1/2 fermions that couple to the gravitini were analyzed in \cite{Bianchi:2000sm}.  It is straightforward for us to do the same for the fermions we study here.

In \cite{Freedman:1999gk, DeWolfe:2000xi, Arutyunov:2000rq, Bianchi:2000sm} fluctuation equations were solved subject to a regularity condition in the deep interior.  The solutions were hypergeometrics characterized by a parameter $a$, defined in our notation as
\eqn{aDef}{
a \equiv -{1 \over 2} + {1 \over 2} \sqrt{1 - {L^4 p^2 \over Q_1^2}} \,,
}
where $p^2 \equiv \omega^2 - k^2$.  The hypergeometric solutions led to two-point functions with a mass gap at
\eqn{MassGapOne}{
\Delta_1 \equiv {Q_1 \over L^2} \,,
}
and a continuous distribution above the gap, appropriate for a state of ${\cal N}=4$ Super-Yang-Mills on the Coulomb branch.

We may perform the analogous analysis for the uncoupled spin-1/2 fields.  Since the gauge fields are zero, the equation cares only about the mass parameters $(m_1, m_2)$, which for the moment we leave arbitrary.  We find for the solution  regular at the deep interior (reverting to $\chi$ using \eno{ChiToPsi}),
\eqn{}{
\chi = (1-v)^{{3 \over 2} + m_1 + m_2} v^{a_{m_1} - {1 \over 6}} F\left({3 \over 2} + m_1 + a_{m_1}, 1 + m_1 + 2 m_2 + a_{m_1}, 2+2a_{m_1}, v \right) \,,
}
where the variable $v$ is given in \eno{vDef}, and we have defined 
\eqn{}{
a_{m_1} \equiv -{1 \over 2} + {1 \over 2} \sqrt{4 m_1^2 - {L^4 p^2 \over Q_1^2}} \,.
}
For all our uncoupled fermions, we have $|m_1| = {1 \over 2}$, and hence $a_{m_1} \to a$ as given in \eno{aDef}.  For all the uncoupled fields in the ${\bf 20}$, both $(m_1, m_2) = (1/2, -1/4)$ and $(m_1, m_2) = (-1/2, 3/4)$,
we find
\eqn{}{
\chi_{\bf 20} = (1-v)^{7/4} v^{a -1/6} F\left(1+a, 2+a, 2+2a,  v \right) \,,
}
while for the fermions in the ${\bf 4}$, with $(m_1, m_2) = (1/2, 1/4)$, we have
\eqn{}{
\chi_{\bf 4} = (1-v)^{9/4} v^{a -1/6} F\left(2+a, 2+a, 2+2a,  v \right) \,,
}
The $\chi_{\bf 20}$ result {\em exactly} matches the positive-chirality solution found in the sector mixed with the gravitino in equation~(188) of \cite{Bianchi:2000sm}; the $\chi_{\bf 4}$ solution has the same 
mass gap \eno{MassGapOne} and continuous spectrum as well.  This completes the analysis of \cite{Bianchi:2000sm} to the entire fermionic spectrum.

We note that for a general $m_1$, the solution would manifest a different gap than found for all the other modes in \cite{Freedman:1999gk, DeWolfe:2000xi, Arutyunov:2000rq, Bianchi:2000sm}.  The universal spectrum is presumably a consequence of the underlying supersymmetry, which constrains the possible values of $m_1$.

\subsection{Conductivities in extremal backgrounds}
\label{sec:GaugePerturbations}

Conductivity studies complement calculations of holographic Fermi surfaces, as both should be interpretable in terms of ${\cal N}=4$ super-Yang-Mills theory.  Here we explain how to formulate bulk differential equations which determine the conductivities.  We will give explicit closed form results for the Coulomb branch solution, defined through the limit \eno{CoulombBranchDefined}.  The nearby extremal 2+1-charge black holes are the ones which exhibit holographic Fermi surfaces which are close to marginal Fermi liquids, and we will provide some numerical results on their conductivities as well.

In order to study conductivities at zero wave-number but non-zero frequency, we may turn on gauge fields at linear order:
 \eqn{aAForm}{
  a_x = e^{-i\omega t} b_x(r) \qquad\qquad
  A_x = e^{-i\omega t} B_x(r) \,.
 }
In general it is necessary at the same time to perturb the metric at linear order:
 \eqn{hForm}{
  g_{tx} = e^{-i\omega t} h_{tx}(r) \,.
 }
Other components of the gauge field may be set to zero as a gauge choice, and other perturbations may be decoupled at linear order from the ones indicated in \eno{aAForm} and \eno{hForm}.  The perturbed Einstein equations reduce to
 \eqn{EEpert}{
  h_{tx}' - 2 A' h_{tx} = 4 X^8 \Phi_1' b_x + {8 \over X^4} \Phi_2' B_x \,,
 }
where $X \equiv e^{-{\phi \over 2\sqrt{6}}}$.  The equation \eno{EEpert} may be used to eliminate $h_{tx}$ from the perturbed Maxwell equations, with the results
 \eqn{aAeoms}{
  b_x'' + \left( 2A' - B' + {h' \over h} + {8 X' \over X} \right) b_x' + 
    {e^{2B} \omega^2 - 4 h X^8 \Phi_1'^2 \over e^{2A} h^2} b_x -
    {8 \Phi_1' \Phi_2' \over e^{2A} h X^4} B_x &= 0  \cr
  B_x'' + \left( 2A' - B' + {h' \over h} - {4 X' \over X} \right) B_x' - 
    {4 X^8 \Phi_1' \Phi_2' \over e^{2A} h} b_x +
    {e^{2B} \omega^2 - 8 h \Phi_2'^2 / X^4 \over e^{2A} h^2} B_x &= 0 \,.
 }
An immediate consequence of these equations is that the flux
 \eqn{ConservedFlux}{
  {\cal F} = {i \over 2} e^{2A-B} h \left( X^8 b_x^* \overleftrightarrow{\partial_r} b_x + 
    {2 \over X^4} B_x^* \overleftrightarrow{\partial_r} B_x \right)
 }
is conserved, in the sense that $\partial_r {\cal F} = 0$.  It is straightforward to show that ${\cal F}$ is the imaginary part of the on-shell action, up to an uninteresting prefactor related to the gravitational constant.  Thus it can be used to compute the real part of the conductivities without extracting the full two-point functions of the currents dual to $A_\mu$ and $a_\mu$.  In particular, if we define
 \eqn{AaDef}{
  {\cal A}_\alpha = \lim_{r \to \infty} (b_x,B_x) \,,
 }
then the Green's functions of the dual currents $j_x$ and $J_x$ satisfies
 \eqn{Frelation}{
  {\cal F} = {\cal A}_\alpha^* (\Im G^R_{xx,\alpha\beta}) {\cal A}_\beta \,,
 }
where by $\Im G^R$ we mean ${1 \over 2i} (G^R - G^{R\dagger})$, and in defining $G^R$ we insist upon purely infalling boundary conditions at the horizon.  The matrix of conductivities satisfies
 \eqn{CondMat}{
  \Re \sigma_{\alpha\beta}(\omega) = \Im {G^R_{xx,\alpha\beta}(\omega) \over \omega} \,,
 }
where by $\Re\sigma$ we mean ${1 \over 2} (\sigma + \sigma^\dagger)$.  Because $\Re\sigma$
is hermitian, it is diagonalizable.  Positivity properties of the Green's function guarantee that these eigenvalues are positive.  The imaginary part of the conductivity, corresponding to the real part of $G^R_{\alpha\beta}$, has an additive ambiguity, linear in $\omega$, which can only be resolved through a suitable holographic renormalization prescription.  Our approach of side-stepping this difficulty by using a conserved flux to compute only the imaginary part of the Green's function is similar to the one in \cite{Gubser:2008sz}; however, we will see that some useful information about the real part of the Green's function is available without the full technology of holographic renormalization.

\subsection{Conductivities for the Coulomb branch solution}
\label{sec:ConductivityCoulomb}

In the Coulomb branch limit defined in \eno{CoulombBranchDefined}, the equations of motion for $h_{tx}$, $b_x$, and $B_x$ all decouple from one another.  The $b_x$ and $B_x$ equations can be simplified to
 \eqn{bBCoulomb}{
  b_x'' + {3r^2-Q_1^2 \over r^3+Q_1^2 r} b_x' + 
    {\omega^2 L^4 \over r^4+Q_1^2 r^2} b_x &= 0  \cr
  B_x'' + {3 \over r} B_x' + {\omega^2 L^4 \over r^4+Q_1^2 r^2} B_x &= 0 \,.
 }
In fact one equation can be transformed into the other by the rescaling
 \eqn{bBrescaling}{
 b_x = X^{-6} B_x \,,
 }
 so the solutions to one may be derived from the other; 
the general solutions to these equations may be expressed in terms of hypergeometric functions.  
These equations were studied in \cite{Bianchi:2000sm}, where it was shown that the solutions are related to the solution of the massless scalar fluctuation;  however, it is useful to review the relevant parts of the calculation from a perspective that will facilitate the numerical work we will do in the next section.  In the rest of this section will be convenient to set $L = Q_1 = 1$, so also for the gap \eno{MassGapOne} $\Delta_1 = 1$. 

In order to determine which solutions are physical, we must examine the behavior near $r=0$, where  the independent solutions to \eno{bBCoulomb} take the form
 \eqn{bBSmallr}{
  \seqalign{\span\TL & \span\TR}{b_x &\sim r^{1 \pm \sqrt{1-(\omega+i\epsilon)^2}}  \cr
  B_x &\sim r^{-1 \pm \sqrt{1-(\omega+i\epsilon)^2}}} \qquad\qquad
     \hbox{for}\quad r \to 0 \,,
 }
where  we have introduced an $i\epsilon$ prescription so that the desired solution corresponds to choosing the plus sign in each exponent, associated to the less singular solution.
The full solutions to \eno{bBCoulomb} are then
 \eqn{bBSolution}{
  b_x &= {\cal B}_1 {\Gamma\left(1 + a_\omega\right)
                   \Gamma\left( 2+a_\omega \right) \over
                   \Gamma\left( 2+2a_\omega \right)}\, 
    r^{2+2a_\omega} \,
    {}_2F_1\left(1+a_\omega,
      2+a_\omega; 2+2a_\omega; -r^2 \right)  \cr
  B_x &= {\cal B}_2{\Gamma\left(1 + a_\omega\right)
                   \Gamma\left( 2+a_\omega \right) \over
                   \Gamma\left( 2+2a_\omega \right)}\, 
    r^{2a_\omega}\,
    {}_2F_1\left( a_\omega,
     1+a_\omega;2+2a_\omega; -r^2 \right) \,,
 }
where the ${\cal B}_\alpha$ are integration constants, and 
\eqn{}{
a_\omega \equiv a(k=0) = -{1\over 2} + {1 \over 2} \sqrt{1 -\omega^2} \,.
}
The gamma functions could have been soaked into the definitions of ${\cal B}_\alpha$, but including them explicitly makes the expansion near the boundary simpler.  In \eno{bBSolution} and below, we mean to continue with the prescription $\omega \to \omega + i\epsilon$, but for simplicity of notation we do not write it explicitly.  Noting that the hypergeometric functions approach unity as $r \to 0$, it is easy to see that
 \eqn{FfrombB}{
  {\cal F} = (|b_x|^2 + 2|B_x|^2) \theta \left(\omega^2-1\right) \sqrt{\omega^2-1} \,.
 }
Using standard identities for hypergeometric functions, one may show that
 \eqn{bBaArelation}{
  {\cal A}_\alpha = M_{\alpha\beta} {\cal B}_\beta \qquad\quad\hbox{where}\qquad\quad
   M_{\alpha\beta} = {\Gamma(2+2a_\omega) \over 
     \Gamma\left( 1+a_\omega\right)
     \Gamma\left( 2+a_\omega \right)}
     \begin{pmatrix} 1 & 0 \\ 0 & 1 \end{pmatrix} \,.
 }
The matrix $M_{\alpha\beta}$ is convenient because it allows us to quickly extract the real part of the conductivities from the flux:
 \eqn{GotSigma}{
  \Re\sigma_{\alpha\beta}(\omega) = {1 \over \omega} (M^{-1\dagger})_{\alpha\gamma} 
    {\partial^2 {\cal F} \over \partial {\cal B}_\gamma^* \partial {\cal B}_\delta}
    (M^{-1})_{\delta\beta} \,.
 }
Because the equations of motion \eno{bBCoulomb} decouple, $M_{\alpha\beta}$ and therefore $\Re\sigma_{\alpha\beta}(\omega)$ are diagonal.  The conductivity eigenvalues are found to be
 \eqn{ReSigmas}{
  \Re \sigma_a(\omega) = {1 \over 2} \Re \sigma_A(\omega) =  {\pi \omega \over 2} 
    \theta(\omega^2-1) \tanh {\pi \sqrt{\omega^2-1} \over 2} 
      \qquad\hbox{for $\omega \neq 0$\,.}
 }
Again we see the structure of a continuum above a gap; this is characteristic of both insulators and hard-gapped superconductors.  The difference between the two, from the point of view of conductivities, is that superconductors have an additional contribution to $\Re \sigma_a(\omega)$ proportional to $\delta(\omega)$ (infinite DC conductivity), whereas insulators do not.

We expect that the Coulomb branch state is an insulator with respect to $A_\mu$ and a superconductor with respect to $a_\mu$.  The reason is that, as demonstrated in \cite{Kraus:1998hv} (and further discussed in \cite{Freedman:1999gk}), the Coulomb branch state under consideration consists of a uniform disk of D3-branes, spread out in the $X_1$ and $X_2$ directions, but localized to the plane $X_3 = X_4 = X_5 = X_6 = 0$, where $X_i$ are the six real scalars of ${\cal N}=4$ super-Yang-Mills.  Rotations in the $X_1$-$X_2$ directions correspond to the $U(1)_1$ gauge symmetry of $a_\mu$ in supergravity, while simultaneous rotations in the $X_3$-$X_4$ and $X_5$-$X_6$ directions correspond to $U(1)_2$ and $A_\mu$.  $U(1)_2$ is genuinely a symmetry of the Coulomb branch state because each of the D3-branes lies at a fixed point of the corresponding rotations in ten dimensions.  $U(1)_1$ is also a symmetry, but only insofar as we regard the disk of branes as truly uniform.  At finite $N$, assuming individual D3-branes must take definite positions within the disk, $U(1)_1$ is explicitly broken.  Even in the formal limit of infinite $N$, corresponding to calculations in classical supergravity, it is reasonable to expect that the main features of an explicitly broken $U(1)_1$ symmetry will arise, namely infinite DC conductivity and the Meissner effect.

The most efficient way to check the expectations explained in the previous paragraph is to use the wave-functions \eno{bBSolution} to extract enough of the Green's function to see the infinite DC conductivity.  To this end, consider the expansions
 \eqn{bBSeries}{
  b_x &= {\cal B}_1 \left[ \left( 1 - {1 \over r^2} \right) + 
   {1 + 2 \log r \over 4r^2} \omega^2 + {\cal O}(\omega^4) + {\cal O}(1/r^4) \right] \cr
  B_x &= {\cal B}_2 \left[ 1 + {1 + 2 \log r \over 4r^2} \omega^2 + {\cal O}(\omega^4) 
     \right] \,.
 }
Formally, we may compute the Green's functions through the expression
 \eqn{FullGreen}{
  \lim_{r \to \infty} {\cal F}^{\bf C} = {\cal A}_\alpha^* G^R_{xx,\alpha\beta} {\cal A}_\beta
 }
where
 \eqn{Fcomplex}{
  {\cal F}^{\bf C} \equiv -e^{2A-B} \left( X^8 b_x^* \partial_r b_x + 
    {2 \over X^4} B_x^* \partial_r B_x \right) \,.
 }
The expression \eno{FullGreen} essentially follows from considerations of the on-shell action, i.e.~the standard AdS/CFT Green's function prescription \cite{Gubser:1998bc,Witten:1998qj}; see \cite{Son:2002sd,Gubser:2008sz}.  But it is a formal expression due to the ${\log r \over r^2}$ in the expansions \eno{bBSeries}, which gives rise to a logarithmic divergence proportional to $\omega^2$ in the limit on the left hand side of \eno{FullGreen}.  In holographic renormalization \cite{Bianchi:2001kw}, this divergence is canceled through the introduction of a boundary counterterm proportional to $f_{mn}^2 + 2 F_{mn}^2$.  The counterterm only effects the real part of the Green's function, and only at quadratic order in $\omega$.  Thus we do not need to establish a definite renormalization prescription if our aim is to read off the $\omega$-independent term.  In short, by plugging \eno{bBSeries} into \eno{FullGreen} and ignoring the divergence, we obtain
 \eqn{ZeroGreen}{
  G^R_{xx,a} = -2 + {\cal O}(\omega^2) \qquad
  G^R_{xx,A} = {\cal O}(\omega^2) \qquad\hbox{at small $\omega$ and $\vec{k}=0$.}
 }
 Recalling the familiar relation,
 \eqn{}{
 \sigma(\omega \approx 0)= {G^R(0) \over i (\omega + i \epsilon)} = G^R(0) \left(-i {\cal P} {1 \over \omega} - \pi \delta(\omega) \right) \,,
 }
and putting \eno{ZeroGreen} together with \eno{ReSigmas}, we arrive at
 \eqn{FullSigma}{
  \Re \sigma_a(\omega) &= 2\pi\delta(\omega) + {\pi \omega \over 2} 
    \theta(\omega^2-1) \tanh {\pi \sqrt{\omega^2-1} \over 2}  \cr
  \Re \sigma_A(\omega) &= \pi \omega \,
    \theta(\omega^2-1) \tanh {\pi \sqrt{\omega^2-1} \over 2} \,,
 }
confirming that the Coulomb branch solution has infinite DC conductivity for the current dual to $a_\mu$ and zero DC conductivity for the current dual to $A_\mu$.

To understand the Meissner effect, we first recall that the two-point function for a conserved current must take the form
 \eqn{GreenCurrent}{
  G_{mn}(k) = G(k^2) \left( g_{mn} - {k_m k_n \over k^2} \right) \,,
 }
where $k^2 = g^{mn} k_m k_n$ and we choose mostly plus signature for the boundary metric $g_{mn}$, which we assume to be the flat Minkowski metric.  Putting \eno{ZeroGreen} and \eno{GreenCurrent} together, we find
 \eqn{LimG}{
  G^R_a(0) = -2 \qquad\qquad G^R_A(0) = 0 \,.
 }
We can now check that there is a Meissner effect for $a_\mu$ and not $A_\mu$, using the London equation.  Following arguments of \cite{Weinberg:1986cq} (see also the discussion in \cite{PufuThesis}), one finds at long wavelengths that $\vec{j}(0,\vec{k}) = G^R_a(0) \vec{a}(0,\vec{k})$, or in position space $\vec{j}(\vec{x}) = G^R_a(0) \vec{a}(\vec{x})$, provided we make the gauge choice $\vec\nabla \cdot \vec{a} = 0$.  Assuming also the Maxwell equation $\nabla \times \vec{B}_a = \vec{j}$ for the magnetic field $\vec{B}_a = \nabla \times \vec{a}$, we arrive at
 \eqn{Meissner}{
  \left[ \nabla^2 + G^R_a(0) \right] \vec{a} = 0 \,.
 }
Because $G^R_a(0) < 0$, solutions in the presence of the condensate are exponentially rising in some direction and falling in another.  The resolution is that the magnetic field penetrates into the condensate a characteristic distance $1/\sqrt{-G^R_a(0)}$ and destroys the condensate where it is large.

It may seem there is a loophole in the argument of the preceding paragraph: in computing $G^R_a(0)$ and $G^R_A(0)$, we first set $\vec{k}=0$ and then took $\omega \to 0$.  But in \eno{Meissner} we used $G^R_a(0)$ in a different limit: $\omega=0$ first with $\vec{k}$ made small afterward.  However, due to the four-dimensional Lorentz invariance of the system, fluctuation solutions with nonzero $\vec{k}$ as well as $\omega$ just take the form \eno{bBSolution} with $a_\omega \to a$, depending only on the combination $p^2 \equiv \omega^2 - k^2$.   The limit is then smooth regardless of whether $\omega$ or $\vec{k}$ is taken to zero first.

\subsection{Conductivities in the marginal Fermi liquid regime}
\label{sec:MFLconduct}

We now wish to investigate conductivities of extremal 2+1-dimensional black holes in the regime where marginal Fermi liquid behavior was found, namely large $Q_1/r_H$, corresponding to $\mu_R \to 0$; we may think of this as ``doping" the 1-charge Coulomb branch background with a little $Q_2$.   The equations of motion \eno{aAeoms} are complicated enough that it would be unenlightening to reproduce them here; however recall the key feature that mixing between $a_\mu$ and $A_\mu$ is proportional to $\Phi_1' \Phi_2'$, and so it vanishes in the strict $r_H \to 0$ limit but is non-zero at any finite $r_H$.
For our class of extremal black holes, the coupled equations \eno{aAeoms} may be diagonalized by either of the two choices 
\eqn{bBdiag}{
b_x = \sqrt{2} \mu_R X^{-6} B_x \,, \quad \quad b_x = - {\sqrt{2} \over \mu_R} X^{-6} B_x \,, 
}
where it is clear that near $\mu_R \to 0$ these correspond to the ``mostly $B_x$" and ``mostly $b_x$" solutions, respectively.

Another key difference between the Coulomb branch case and the case of extremal black holes with small but non-zero $r_H$ is that there is an $AdS_2 \times {\bf R}^3$ region close to the horizon.  The radial dependence of the perturbations near the horizon may be expressed as
 \eqn{LeadingDependence}{
  b_x &= {\cal B}_1 e^{i r_H \sqrt{1+r_H^2} \omega \over 2 (r-r_H) (1+4r_H^2)} \;
     (r-r_H)^{-{i (1+3r_H^2+4r_H^4) \omega \over 2 \sqrt{1+r_H^2} (1+4r_H^2)^2}} 
      \left[ 1 + {\cal O}(r-r_H) \right] \,, \cr
  B_x &= {\cal B}_2 e^{i r_H \sqrt{1+r_H^2} \omega \over 2 (r-r_H) (1+4r_H^2)} \;
     (r-r_H)^{-{i (1+3r_H^2+4r_H^4) \omega \over 2 \sqrt{1+r_H^2} (1+4r_H^2)^2}}
      \left[ 1 + {\cal O}(r-r_H) \right] \,.
 }
The leading behavior is the same (up to the integration constants ${\cal B}_\alpha$), but the corrections in square brackets are not.  It is worth noting that the asymptotic forms \eno{LeadingDependence} do {\it not} reduce to solutions of the form \eno{bBSmallr} when $r_H \to 0$.  This is because of order of limits: In \eno{bBSmallr} we set $r_H=0$ first and then took $r \to 0$; here we take $r$ close to the horizon at finite $r_H$ first.  

Conceptually, our strategy is very similar to the one outlined in the previous subsection: We first use \eno{LeadingDependence} to write the flux as
 \eqn{ConservedFluxNum}{
  {\cal F} = \omega \sqrt{1+r_H^2} \left( {1 \over 2 r_H^2} |b_x|^2 + 4 r_H^2 |B_x|^2 \right)
    \,,
 }
and then we relate the coefficients ${\cal B}_\alpha$ to the boundary values ${\cal A}_\alpha$ of the gauge fields in preparation for computing the Green's function.
Our numerical scheme is to start with the asymptotic forms \eno{LeadingDependence}, including explicit expressions for the ${\cal O}(r-r_H)$ corrections, and use them to seed a numerical integration of the equations of motion out to large $r$, where we evaluate ${\cal A}_\alpha$.  

\begin{figure}
\begin{center}
\includegraphics[width=4in]{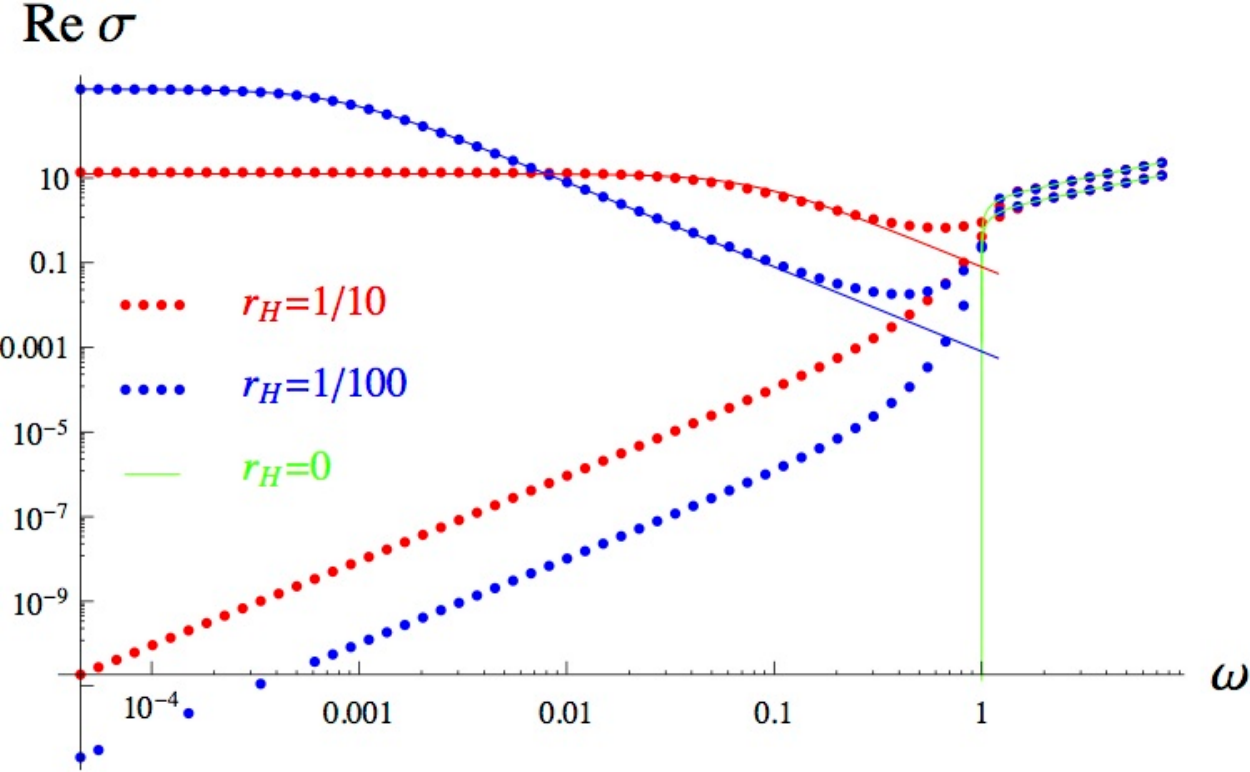}
  \caption{The eigenvalues of the real part of the conductivity matrix for the currents dual to $a_x$ and $A_x$, as functions of $\omega$, for black hole backgrounds close to the Coulomb branch solution.  Red dots show evaluations of the conductivities for $r_H=1/10$; blue dots are for $r_H = 1/100$; and the green curves are the analytic expressions \eno{ReSigmas} for the Coulomb branch solutions.  The blue and red curves are based on the Drude model, as explained in the main text.\label{fig:Conductivities}}
\end{center}
 \end{figure}

The linearity of the equations guarantees that ${\cal A}_\alpha = M_{\alpha\beta} {\cal B}_\beta$ for some $M_{\alpha\beta}$, but because the equations couple, $M_{\alpha\beta}$ is not diagonal. It can however be extracted from numerical solutions for any given $\omega$.  Then we extract the matrix of conductivities precisely as indicated in \eno{GotSigma}, using the near-horizon form \eno{ConservedFluxNum} for the flux.  Finally, we extract conductivity eigenvalues, corresponding to the diagonalization \eno{bBdiag}.  The results are shown in figure~\ref{fig:Conductivities}.

Several points are worth noting:
 \begin{itemize}
  \item As $r_H \to 0$, the mixing between $a_x$ and $A_x$ becomes small, and as a result, the eigenvectors of $\Re\sigma_{\alpha\beta}(\omega)$ are nearly $\begin{pmatrix} 1 \\ 0 \end{pmatrix}$ and $\begin{pmatrix} 0 \\ 1 \end{pmatrix}$, consistent with \eno{bBdiag}.  We will refer to the corresponding conductivity eigenvalues as $\sigma_a(\omega)$ and $\sigma_A(\omega)$, since they predominantly refer to the conductivities of the gauge fields dual to $a_\mu$ and $A_\mu$, respectively.
  \item For $\omega > 1$, $\sigma_a(\omega)$ and $\sigma_A(\omega)$ converge quickly to the results \eno{ReSigmas} found for the Coulomb branch.  In particular, $\sigma_a(\omega) \approx {1 \over 2} \sigma_A(\omega)$.
  \item For $\omega < 1$, $\sigma_A \ll \sigma_a$.  Thus $\sigma_A$ and $\sigma_a$ cross at $\omega \approx 1$.
  \item $\sigma_A \approx \omega^2 r_H^2$ for small $\omega$ and small $r_H$.  For fixed $r_H$, this power law behavior is approximately respected all the way up to $\omega \approx 0.3$.
  \item For small $\omega$ and small $r_H$, one finds
 \eqn{DrudeFit}{
  \Re\sigma_a \approx {\sigma_0 \over 1 + \omega^2 \sigma_0^2} \qquad\hbox{where}\qquad
   \sigma_0 \equiv {1 \over 8r_H^2} \,.
 }
Again the fit is good up to $\omega \approx 0.3$.
 \end{itemize}
The functional form in \eno{DrudeFit} is based on the Drude model, in which more generally
 \eqn{sigmaDrude}{
  \Re \sigma_{\rm Drude} = {\sigma_0 \over 1 + \omega^2 \tau^2} \,,
 }
and, conventionally, $\tau$ is the scattering time while $\sigma_0 = n e^2 \tau / m$ is a constant related to the density of charge carriers.  We note that for small $\sigma_0$, the form \eno{DrudeFit} converges in the sense of distributions to $\sigma_a \approx \pi \delta(\omega)$ at small $\omega$, which is half as large as the Coulomb branch result \eno{FullSigma}.\footnote{In taking this limit, it is important that we are considering $\omega$ to be valued over the entire real line.  Alternatively, if we restrict $\omega\geq 0$, then the distribution $\delta(\omega)$ should be regarded as supported half in the $\omega > 0$ domain and half in the $\omega < 0$ domain.  In this alternative approach we would say $\int_0^\epsilon d\omega \, \pi\delta(\omega) = {\pi \over 2}$, and also $\lim_{\sigma_0 \to 0} \int_0^\epsilon d\omega \, {\sigma_0 \over 1 + \omega^2 \sigma_0^2} = {\pi \over 2}$ where $\epsilon$ stays fixed in the limit.}  This apparent discrepancy is surprising, but it may find its explanation in the existence of an additional $\pi \delta(\omega)$ contribution to $\sigma_a(\omega)$ which our present methods do not allow us to detect.

\section{The 2-charge black hole and soft  gap}
\label{TwoChargeSec}

We turn now to a study of fermion fluctuations in the 2-charge black hole, which has similarities and differences to the 1-charge case; a difference is that here, there is a genuine extremal black hole with a charge density, breaking four-dimensional Lorentz invariance.  As described in section~\ref{ExtremalRelationSec}, like the 1-charge case the limit of the extremal 2+1-charge black holes is not precisely the 2-charge extremal black hole; unlike the 1-charge case, however, we have no fermion fluctuations that ignore the discrepancy.  Therefore the behavior of our fermion fluctuations in the extremal 2-charge solution is disconnected from the 2+1-charge results of section~\ref{TwoPlusOneSec}.  We can nonetheless study them in their own right.  The near-horizon region of the 2-charge black hole is not $AdS_2 \times {\mathbb R}^3$, but is conformal to it \cite{Gubser:2012yb} and the ten-dimensional lift contains an $AdS_3$ factor \cite{Gubser:2009qt}.

\subsection{Fermion eigenvectors in 2-charge background}
\label{TwoChargeEqnsSec}

The 2-charge black hole has $a_\mu = 0$, and hence the charges $q_1$ and $p_1$ are irrelevant.  A number of Dirac equations that are distinct for the 2+1-charge background now coincide.
Below we list the positively-charged modes, both from ``maximal" eigenvectors and the gravitino-free non-maximals:
\begin{center}
\begin{tabular}{|c|c|c|c|c|} \hline
&$m_1$ & $m_2$ & $q_2$  & $p_2$\\ \hline
$2  \chi$, $2  \bar\chi$ & ${1 \over 2}$ & $- {1 \over 4}$ & $2$ & $0$ \\
$3  \chi$, $3  \bar\chi$ & ${1 \over 2}$ & $- {1 \over 4}$& $1$ & $-{1 \over 2}$ \\ 
$1  \chi$, $1  \bar\chi$ & $-{1 \over 2}$ & ${3 \over 4}$ & $1$ & ${1 \over 2}$ \\
$  1\chi$, $ 1 \bar\chi$ & ${1 \over 2}$ & $ {1 \over 4}$ & $1$ & $-{1 \over 2}$ \\ \hline
\end{tabular}
\end{center}
The first column counts the number of modes of each chirality with the corresponding set of charges.
We note that each charge is chirally balanced by itself, unlike in the 2+1-charge case; this reflects the fact that $U(1)_2$ is non-anomalous.  

There are also negatively-charged conjugate modes with opposite $q_2$ and $p_2$.  Finally, there are  neutral modes, with balanced chiralities and two different mass functions:
\begin{center}
\begin{tabular}{|c|c|c|c|c|} \hline
&$m_1$ & $m_2$ & $q_2$  & $p_2$\\ \hline
$2  \chi$, $2  \bar\chi$ & $-{1 \over 2}$ & ${3 \over 4}$ & $0$ & $0$ \\ 
$2  \chi$, $2  \bar\chi$ & ${1 \over 2}$ & $ {1 \over 4}$& $0$ & $0$ \\ \hline
\end{tabular}
\end{center}

\subsection{Near-horizon analysis: extremal 2-charge case}

We can perform a general near-horizon analysis of the Dirac equation \eno{DiracEqn} in the extremal 2-charge case. Near the horizon we have
\eqn{}{
e^{2A} \to {r^{2/3} Q_2^{4/3} \over L^2} \,, \quad\quad e^{2B} \to {r^{2/3} L^2 \over Q_2^{8/3}} \,, \quad \quad h \to {2 r^2 \over Q_2^2} \,,  \quad \quad
\Phi_2 \to - { r^2 \over 2 Q_2 L } \,,
}
while for the scalar, 
\eqn{}{
\phi \to \sqrt{2 \over 3} \log \left( Q_2^2 \over r^2\right) \,, \quad \to \quad
e^{\phi \over \sqrt{6}} \to {Q_2^{2/3} \over r^{2/3}} \,.
}
Like the 1-charge case but unlike the (2+1)-charge cases, the scalar diverges at the horizon.  The mass term in the limit becomes
\eqn{}{
m(\phi) \to {2 m_2 Q_2^{4/3} \over L r^{4/3}} \,,
}
which is also divergent.  
Looking at the second order equations, we find
\eqn{}{
\partial_r^2 \Psi_{\alpha \pm} + {2 \over r} \partial_r \Psi_{\alpha \pm}  +{ \omega^2 L^2 - 8 m_2^2 Q_2^2  \over 4 r^4} \Psi_{\alpha \pm} = 0\,.
}
The two dominant contributions to the zero-derivative term are the frequency and the diverging mass.  For $\omega \neq 0$ this has the solutions
\eqn{TwoChargeSolns}{
\Psi \sim  \exp \left[\pm{1 \over 2r} \sqrt{8 m_2^2 Q_2^2 - \omega^2 L^4} \right] \,.
}
for either $\Psi_{\alpha + }$ or $\Psi_{\alpha -}$.  We notice that for infinitesimal $\omega$, these solutions are not propagating waves; the divergence of the scalar has created a certain minimal value of $\omega$ necessary to make an infalling solution:
\eqn{DeltaTwo}{
\Delta_2 = {2  \sqrt{2} |m_2| Q_2 \over L^2} \,.
}
We can also study the $\omega = 0$ equation directly, and then take the small $r$ limit,  The different order of limits changes the coefficient of the first derivative term:
\eqn{TwoChargeZeroOmega}{
\partial_r^2 \Psi_{\alpha \pm} + {1 \over r} \partial_r \Psi_{\alpha \pm}  - {2 m_2^2 Q_2^2 \over r^4} \Psi_{\alpha \pm} = 0\,.
}
Notice how this equation differs from the analogous case for the 2+1-charge black holes \eno{NearHorizonDirac} in having a leading $1/r^4$ term generated by the diverging mass, dominating over the $1/r^2$ term that would contain $\nu_k^2$.
Equation \eno{TwoChargeZeroOmega}  has modified Bessel function solutions
\eqn{BesselSolns}{
\Psi \sim K_0\left(\sqrt{2 m_2^2 Q_2^2} \over r \right), \; I_0\left(\sqrt{2 m_2^2 Q_2^2} \over r \right)
}
which for small $r$ have the behaviors
\eqn{BesselSolnsLimit}{
\Psi \sim r^{1/2} \exp \left[\pm{1 \over r} \sqrt{2 m_2^2 Q_2^2 } \right] \,,
}
similar to the finite-$\omega$ solutions \eno{TwoChargeSolns} with an additional $r^{1/2}$ factor.

The  behavior brought on by the diverging mass in these known ${\cal N}=8$ gauged supergravity modes should be compared to \cite{Gubser:2012yb}, where Dirac equations with constant mass were studied in the extremal 2-charge black hole; there no analog to \eno{DeltaTwo} was found.  This is a case in which the top-down example is instructive, as the simpler  solutions of \cite{Gubser:2012yb} are apparently not present in any of the actual modes known to be dual to operators ${\cal N}=4$ Super-Yang-Mills; it would be interesting to diagonalize the gravitino sector to see if the  behavior in the solutions of \cite{Gubser:2012yb} is present there.

\subsection{Fermi surfaces for 2-charge black hole}
\label{sec:FS2Q}

We numerically solved the Dirac equations listed in section~\ref{TwoChargeEqnsSec} at $\omega = 0$, subject to the regular boundary condition at the horizon, which is the $K_0$ Bessel function in \eno{BesselSolns}, leading to the minus sign in \eno{BesselSolnsLimit}.  As in the 2+1-charge case, we solved the equations out to the boundary and looked for values of $k = k_F$ such that $A(k_F)$ vanishes.

For the first two equations in the table, both with mass values $(m_1, m_2) = (1/2, -1/4)$, we found such Fermi-surface-type singularities.  In the first case, that with $(q_2, p_2) = (2, 0)$ the corresponding value of $k_F$ is
\eqn{}{
\frac{k_F}{\mu_2} = 0.83934 \,,
}
while for the second case of $(q_2, p_2) = (1, -1/2)$ we obtain
\eqn{}{
\frac{k_F}{\mu_2} = 0.05202 \,.
}
We find no Fermi surface singularities for the remaining equations.
Unlike the (2+1)-charge case, there is no analogue of $\nu_k$ and no analogous excitations;
the form of the fluctuations \eno{TwoChargeSolns}
prevents modes with small $\omega$ and $k_\perp = k - k_F$ from manifesting an ordinary infalling mode. To find such a solution  one must go to $\omega \geq \Delta_2$, where the small-$\omega$ form of the Green's function \eno{GreensFunction} no longer obtains.  
A more thorough understanding of these fluctuations is naturally of interest, and we leave it to future work.

\subsection{Conductivities for the 2-charge black hole}

\begin{figure}
\begin{center}
\includegraphics[width=4in]{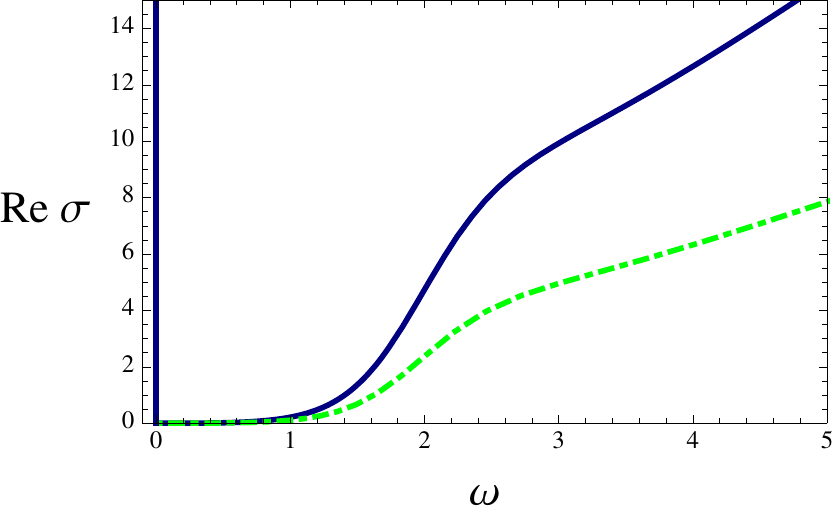}
  \caption{The real part of the conductivity  for the currents dual to $b_x$ and $B_x$, as functions of $\omega$, for the extremal 2-charge black hole.  The blue (solid) curve corresponds to the $B_x$ mode, and has a delta function contribution at zero frequency. The green (dot-dashed) line is for the current dual to $b_x$, and the DC conductivity is insulated. Note the presence of a soft gap in each.\label{fig:2QCond}}
\end{center}
 \end{figure}

One may also consider the gauge field fluctuations and associated conductivities.  As in the 1-charge background, the $b_x$ and $B_x$ fluctuation equations \eno{aAeoms} decouple, and one can be transformed into the other by \eno{bBrescaling}.  While we cannot solve the equations analytically for general $\omega$, the $\omega = 0$ solutions are easy to obtain (the $B_x$ case was previously found in \cite{DeWolfe:2011ts}).  The zero frequency equations of motion are
\eqn{bB2Charge}{
  b_x'' + {10+9r^2+3r^4\over (r+r^3)(2+r^2)}\, b_x' + 
     &= 0 \,, \cr
  B_x'' + \left({1 \over r}+{2r\over 2+r^2}\right) B_x' - {8 \over (2+r^2)(r+r^3)^2} B_x &= 0 \,,
 }
 where we have set $Q_2 = L = 1$,
and the solutions regular at the horizon look like
\eqn{reg2Q}{
b_x = {\cal B}_1 \,, \quad \quad B_x = {\cal B}_2 {r^2 \over Q_2^2 + r^2} \,.
}
In the spirit of \eno{Fcomplex}, we can formally define the complex fluxes from \eno{bBCoulomb}, in this case
\eqn{2QF}{
\mathcal{F}^{\bf C}=-h \,e^{2A-B}\left(X^8\, b_x^*\partial_r b_x +\frac{2}{X^4}B_x^*\partial_r B_x\right) \,.
}
 Inserting the regular solutions \eno{reg2Q} into \eno{FullGreen} via the flux \eno{2QF}, we thus extract the zero frequency contributions to the Green's function :
 \eqn{}{
 G_{b}^R(0) = 0 \,, \qquad\qquad G_B^R(0)=-4 \,.
 }
 Accordingly, the real part of the conductivity for the current dual to $B_x$ has a delta function contribution $4\pi\delta(\omega)$, while the current dual to $b_x$ does not. Thus in both the 1-charge and 2-charge cases, it is the gauge field turned on in that class of backgrounds whose dual current manifests infinite conductivity at zero frequency.  Here, however, we have a nonzero charge density \eno{TwoThermo}, and lack the four-dimensional Lorentz invariance that we used in the 1-charge case to demonstrate a Meissner effect.  Thus despite the complementary appearance of the two examples, a more careful investigation is necessary to find out whether the delta function is entirely a consequence of translation invariance \cite{Hartnoll:2008vx}, or if there really is superconductivity.\footnote{We thank C.~Herzog for a discussion on this point.}

While we cannot analytically solve the fluctuation equations \eno{bB2Charge} for nonzero $\omega$, we are able to do so numerically.  In figure~\ref{fig:2QCond}, the real part of the optical conductivity for both currents is plotted as a function of frequency. The most notable feature in this picture is the appearance of a ``soft" gap at $\Delta_{\sigma}=2Q_2/L^2$. This is to be contrasted with the ``hard" gap characteristic of the 1-charge Coulomb branch conductivities \eno{ReSigmas}. Interestingly, this gap appears to be larger than the scale \eno{DeltaTwo} felt by the fluctuations about the 2-charge black hole  Fermi surfaces  by a factor $\Delta_{\sigma}/\Delta_{2} = 2\sqrt{2}$.

\section{Conclusions}
\label{ConclusionsSec}

We have shown that Fermi surface behavior is ubiquitous in strongly coupled ${\cal N}=4$ Super-Yang-Mills theory at zero temperature and finite density; indeed for every realized value of the chemical potentials there are multiple fermionic modes manifesting a Fermi surface singularity.  These Fermi surfaces are non-Fermi liquids for generic backgrounds; indeed they are often in the deep non-Fermi regime, with excitations characterized by large dispersion relation exponents, as well as small width/energy ratios.

There are exceptions to this general behavior.  In particular, one fermion approaches the marginal Fermi liquid limit, as well as vanishing zero-temperature entropy.  This limit is controlled by the 1-charge Coulomb branch solution; both this and the 2-charge extremal black hole (which also has zero entropy at zero temperature) have the conductivity associated to the active gauge field infinite at zero frequency, while the other is insulating; the gaps are hard and soft, respectively.

There are a number of open questions, many related to the dual field theory, which thanks to the top-down construction one may explore explicitly.  One issue is the nature of the quasiparticle-like excitations forming the Fermi surfaces; in some top-down models arguments that have been made for a ``mesino" --- a gauge singlet scalar/fermion bound state (see for example \cite{Huijse:2011hp,Iqbal:2011ae,Huijse:2011ef}).  In \cite{DeWolfe:2011aa}, it was pointed out that the manifest $N^2$ scaling of the top-down Green's functions suggests a particle in the adjoint, coinciding with a suggestion \cite{Gubser:2009qt} that the gaugino itself generated the Fermi surface.  Here we see, however, that operators containing the same gaugino but different scalars (such as {\em e.g.}~Tr $\lambda_1 Z_2$ and Tr $\lambda_1 Z_1$, case 1 and case 2 respectively) do not have the same Fermi surface singularities at values of $\mu_R$ away from the 3-charge black hole point.  Thus the scalars are clearly important; along with the $N^2$ scaling this suggests that adjoint scalar/fermion bound states may be relevant.  (Operators like Tr $F_+ \lambda_1$ are never found to have Fermi surfaces.)
A Luttinger count of the charge density generalizing \cite{DeWolfe:2011aa} (see also \cite{Hartnoll:2010xj, 
Huijse:2011hp, Huijse:2011ef, Hartnoll:2011fn, Iqbal:2011bf,  Hashimoto:2012ti}) could be helpful.

Another interesting question is the role of thermodynamic and superconducting instabilities.  A number of such instabilities are known to be present at nonzero temperature in the family of geometries we discuss (see for example \cite{Gubser:2009qt, Donos:2011ff, Gentle:2012rg}).  It would be valuable to understand better the effects of these instabilities, and whether they vanish in, for example, the marginal Fermi liquid limit.

These calculations can be generalized in a number of obvious ways.  All three chemical potentials of the $SO(6)$ R-symmetry could be allowed to vary independently; we expect such a generic black hole will be similar to our 2+1-charge cases, as the 3-charge case is.  We have also not diagonalized the gravitino mixing;  in the 1-charge Coulomb branch case \cite{Bianchi:2000sm} the gravitino sector led to very similar results as the uncoupled fermions, but it is possible in the general case new phenomena may lurk there.  One could also extend this analysis to the three-dimensional ABJM case, where one mode was already studied in \cite{DeWolfe:2011aa}, or to the six-dimensional $(2, 0)$ theory.

Because we find a marginal Fermi liquid and small zero-temperature entropy in the limit where we approach the one-charge Coulomb branch solution, it is natural to inquire how closely our construction might come to what is needed to describe important strongly coupled condensed matter systems, such as cuprates near optimal doping and heavy fermion compounds.  We have a top-down construction, so we are finally in a position to formulate questions about a known strongly coupled field theory (namely ${\cal N}=4$ super-Yang-Mills) whose answers, supergravity tells us, include the existence of a marginal Fermi liquid.  An important feature of our Coulomb branch construction, however, is the existence of a condensate of the adjoint scalar fields in ${\cal N}=4$ super-Yang-Mills, and these are fundamental scalars, not fermion composites.  The condensate breaks $SU(N)$ down to $U(1)^N$, and the ${\cal O}(N)$ modes associated with the unbroken $U(1)^N$ are massless because they correspond to motions of individual D3-branes in the directions transverse to the world-volume---which is to say, directions in which the presence of the D3-branes spontaneously breaks translational symmetries.  One can also argue that these modes are massless simply because of the unbroken gauge invariance associated with them; at any rate, first-principles field-theory arguments show that there are ${\cal O}(N)$ massless modes.  Remarkably, we learn from supergravity that the ${\cal O}(N^2)$ non-abelian degrees of freedom are all gapped in the strict Coulomb branch limit when $g_{YM}^2 N$ is large: the evidence for this is the gap in the spectral measure observed, for example, in \eno{FullSigma}.  As we depart slightly from the Coulomb branch limit, we observe the marginal Fermi liquid behavior at low energies and small $k_F$, as compared to the energy and momentum scale set by the gap.  It is reasonable to suppose that the ${\cal N}=4$ super-Yang-Mills dynamics above the gap can be integrated out, leading to some effective theory that controls the marginal Fermi liquid behavior that we see.  It would be very instructive to develop field theory arguments that support this picture without explicit reference to supergravity.  It may happen that at subleading order in $N$, the ${\cal O}(N)$ massless modes give rise to interactions that smooth out the Fermi surface singularity.  Even if the specific construction we have laid out is not directly related to specific condensed matter systems, a first-principles understanding of how strongly coupled field theory dynamics leads (in some appropriate limit) to a marginal Fermi liquid would clearly be an important advance.

\section*{Acknowledgments}

We are grateful to Tom DeGrand, Sean Hartnoll, Michael Hermele, Christopher Herzog, Shamit Kachru, Subir Sachdev, and Amos Yarom for helpful discussions and correspondence.  We especially thank Shivaji Sondhi for an interesting discussion about possible field theory interpretations of the marginal Fermi liquid regime.  The work of O.D.\ and C.R.\ was supported by the Department of Energy under Grant
No.~DE-FG02-91-ER-40672.  The work of S.S.G.\ was supported by the Department of Energy under Grant No.~DE-FG02-91ER40671.

\bibliographystyle{JHEP}
\bibliography{Fermi}

\end{document}